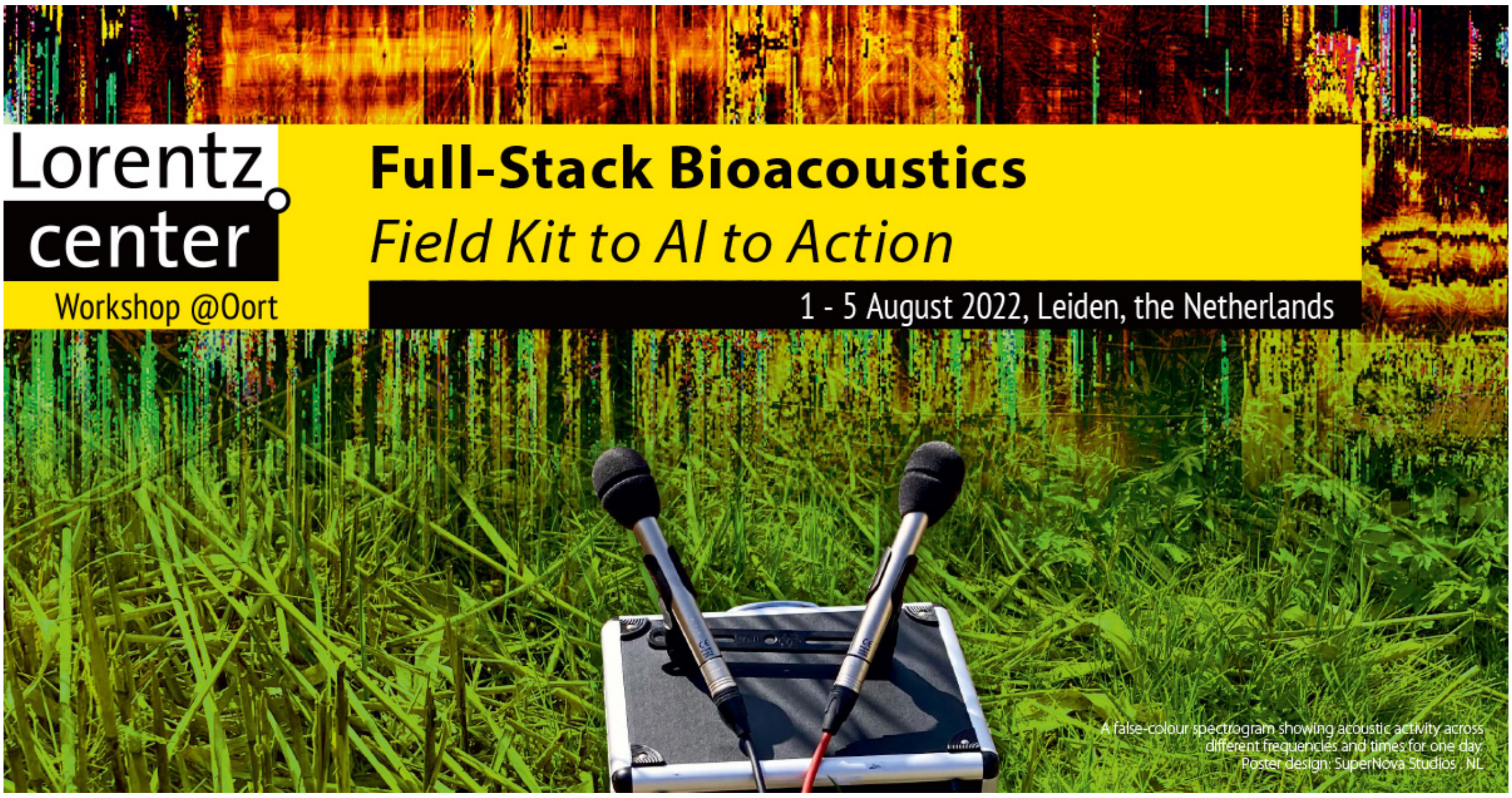

**Workshop report**
**Full-Stack Bioacoustics: Field Kit to AI to Action**
**Lorentz Center, Leiden, the Netherlands**
**1-5 August 2022**

Scientific organisers:

- Dan Stowell, Naturalis Biodiversity Center / Tilburg University
- Caitlin Black, Universiteit van Amsterdam
- Florencia Noriega, CODE University of Applied Sciences Berlin
- Sarab Sethi, University of Cambridge

This report contains an overview of the workshop aims and structure, as well as reports from the six groups.

## Scientific case

Acoustic data (sound recordings) are a vital source of evidence for detecting, counting, and distinguishing wildlife. This domain of "bioacoustics" has grown in the past decade due to the massive advances in signal processing and machine learning, recording devices, and the capacity of data processing and storage. Numerous research papers describe the use of Raspberry Pi or similar devices for acoustic monitoring, and other research papers describe automatic classification of animal sounds by machine learning. But for most ecologists, zoologists, conservationists, the pieces of the puzzle do not come together: the domain is fragmented. In this Lorentz workshop we bridge this gap by bringing together leading exponents of open hardware and open-source software for bioacoustic monitoring and machine learning, as well as ecologists and other field researchers. We share skills while also building a vision for the future development of "bioacoustic AI".

## Design of the workshop programme

Our aim, in general terms, was to use the week ***to make progress in an immature and cross-disciplinary field***, e.g. to facilitate cross-disciplinary discussions, and bring unresolved issues to the surface. The following principles and general structure are given in case useful for events in other fields which are at this same stage of development.

Principles (and plans derived from them):

- **Group-work** is good for interrogating a topic (as opposed to the largely-passive mode of seminars/tutorials/panels).
  - ↳ Around half of the scheduled time was dedicated to groupwork. Heuristically, teams should be no bigger than 7 people to be effectve: we had six groups of approximately 7 people each.
- Participants from **multiple fields need to be able to understand each other**
  - ↳ Plenary lectures on the first day were designed to give an overview, plus other focal talks throughout the week. Additionally, a tutorial session "Everything you wanted to know about [machine learning]/[acoustic surveying]" whose content was based on questions gathered in a pre-survey.
- Groups should contain members with ***complementary* skills & needs**
  - ↳ In a pre-survey we elicited topic keywords from participants (topics "offered" and topics "wanted") and used this to pre-allocate groups to maximise complementarity. (Participants were able to switch group on day 1 during initial goals discussion.)
- Groups need **mutual awareness of each *individual*'s state of thought**, and participants' varying communication styles might not automatically support this.
  - ↳ We used the software developer's concept of the "Daily standup" meeting: a 15 minute daily meeting restricted so that each person answers three questions: "*What did you do yesterday? What will you do today? Anything blocking your progress?*" In practice we found that this did indeed help maintain awareness and facilitate problem-solving in the short time available.
- Participants need **awareness across groups**, and not just within their own group, to ensure they benefit from the insights of the whole cohort
  - ↳ The daily standups were organised in a rota with groups combined pairwise: in other words, each day a group hears a detailed update from one other group.
  - ↳ We also provided collective communication tools (Slack, Notion), in addition to the informal conversation that occurs at an in-person event.
- **A practical task to work on together** helps give focus to the groupwork, and to serve as a "worked example" through which the groups can surface and examine unresolved issues. The task could also be theoretical or analytical.
  - ↳ In our case the suggested template for each group's practical task was to deploy an acoustic monitoring device and then analyse data from it. We prepared acoustic monitoring devices and outdoor surveying sites in advance.

<table>
<tr><th></th><th colspan="2">09:00</th><th colspan="2">10:00</th><th colspan="2">11:00</th><th colspan="2">12:00</th><th colspan="2">13:00</th><th colspan="2">14:00</th><th colspan="2">15:00</th><th colspan="2">16:00</th><th colspan="2">17:00</th><th colspan="2">18:00</th><th colspan="2">19:00</th><th colspan="1">20:00</th></tr>
<tr><td>Mon</td><td>--</td><td>Registration, tea & coffee</td><td>Intro</td><td colspan="4">Plenary state-of-the art/overview lectures (6 x 20 min)</td><td colspan="4">Lunch & informal discussion</td><td>Hardware lightning talks</td><td>Tea & coffee</td><td colspan="5">Practical: Deploying hardware devices into the "field"</td><td colspan="2">Welcome reception</td><td colspan="3"></td></tr>
<tr><td>Tue</td><td colspan="2">Plenary lecture: Sig proc & machine learning</td><td>Tea & coffee and group-formation</td><td colspan="3">Group-work</td><td colspan="4">Lunch & informal discussion</td><td>Plenary lecture</td><td colspan="2">Group-work</td><td>Tea & coffee</td><td colspan="2">Group-work</td><td colspan="7"></td></tr>
<tr><td>Wed</td><td colspan="2">Semi-plenary: "Everything you wanted to know about ___"</td><td>Tea & coffee</td><td>"Stand-up" meeting</td><td colspan="2">Group-work</td><td colspan="3">Lunch & informal discussion</td><td>Open science and data sharing</td><td>Plenary lecture</td><td colspan="2">Group-work</td><td colspan="7">Outing (dunes/lakes)</td><td colspan="3">Workshop dinner</td></tr>
<tr><td>Thu</td><td colspan="2">Plenary lecture: Ecology</td><td>Tea & coffee</td><td>"Stand-up" meeting</td><td colspan="2">Group-work</td><td colspan="4">Lunch & informal discussion</td><td>Plenary lecture</td><td>Initial tutorial paper discussion</td><td>Group-work</td><td>Tea & coffee</td><td colspan="2">Group-work</td><td colspan="7"></td></tr>
<tr><td>Fri</td><td>"Stand-up" meeting</td><td colspan="2">Group-work</td><td>Tea & coffee</td><td colspan="3">Groups report back to all</td><td colspan="3">Lunch & informal discussion</td><td>Collective drafting of final outcomes</td><td colspan="2">Initial drafting of tutorial paper</td><td>Depart</td><td colspan="9"></td></tr>
</table>

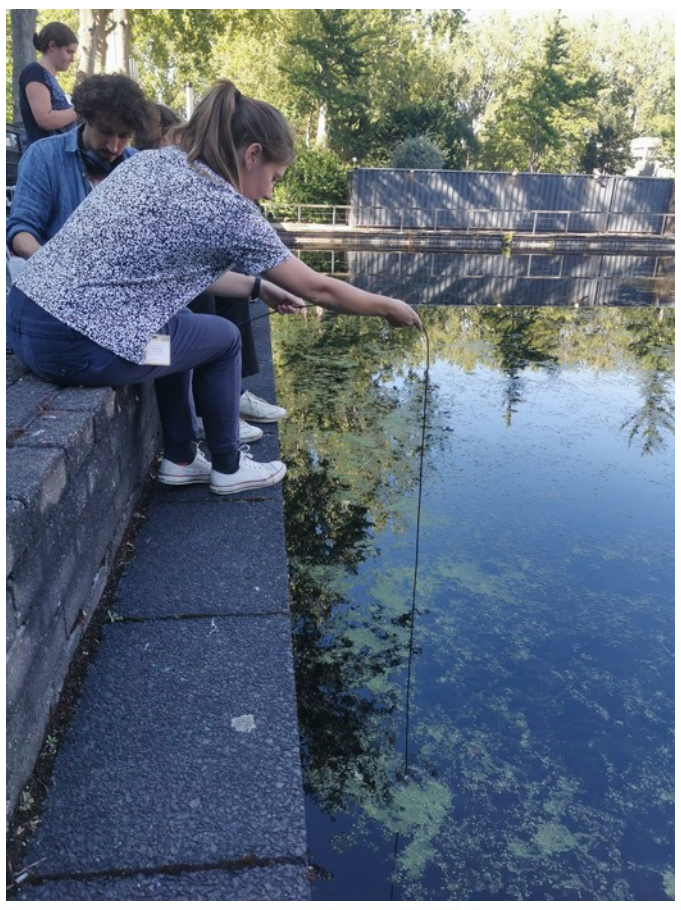
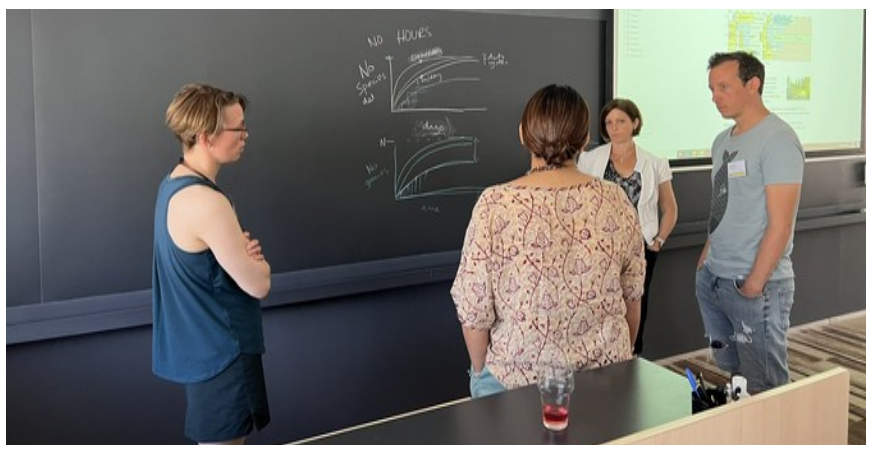
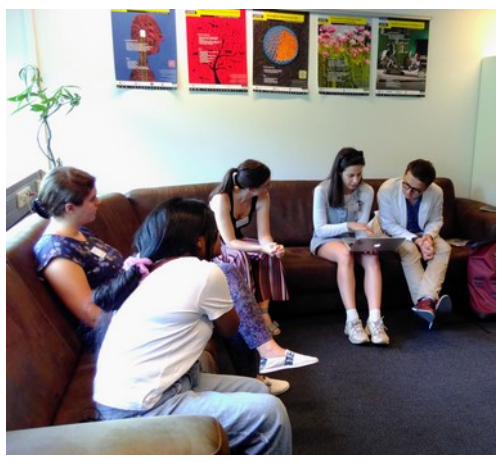
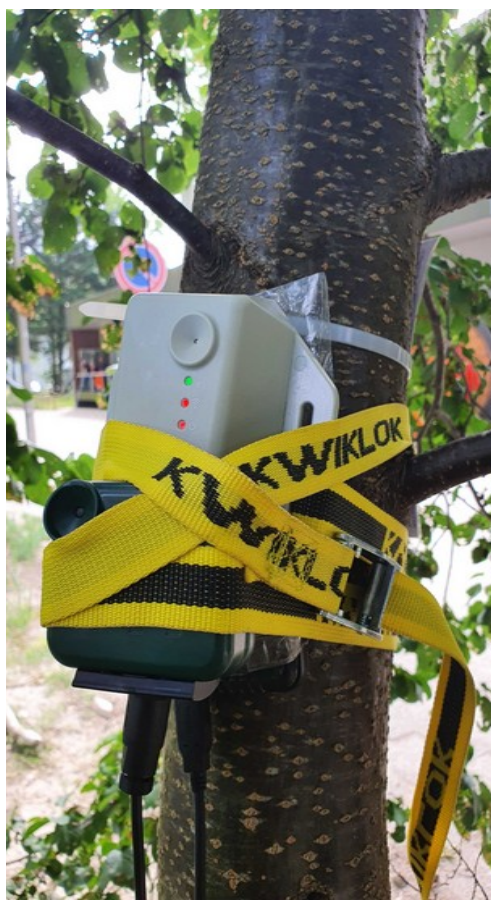

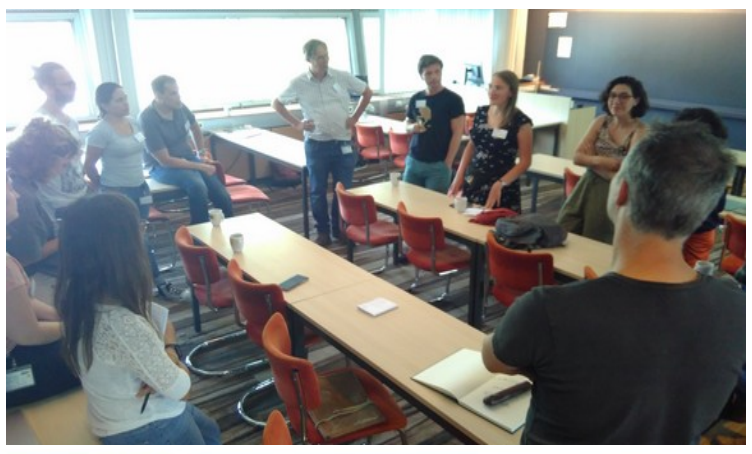
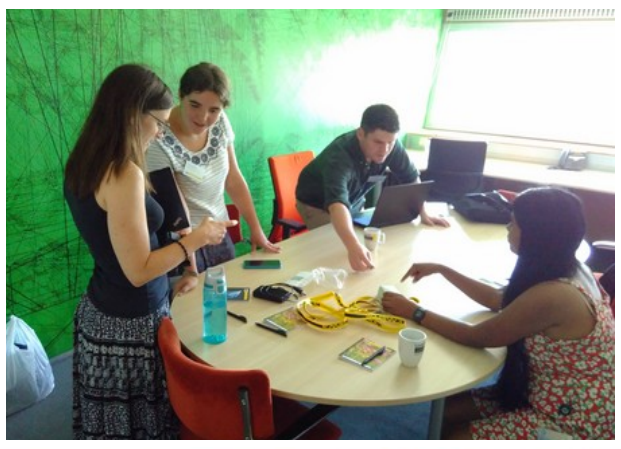
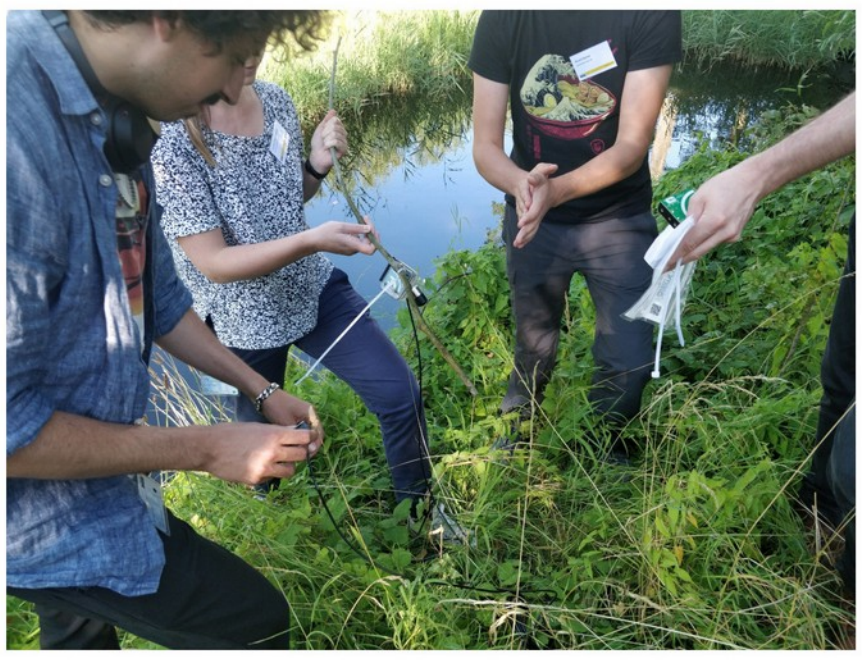
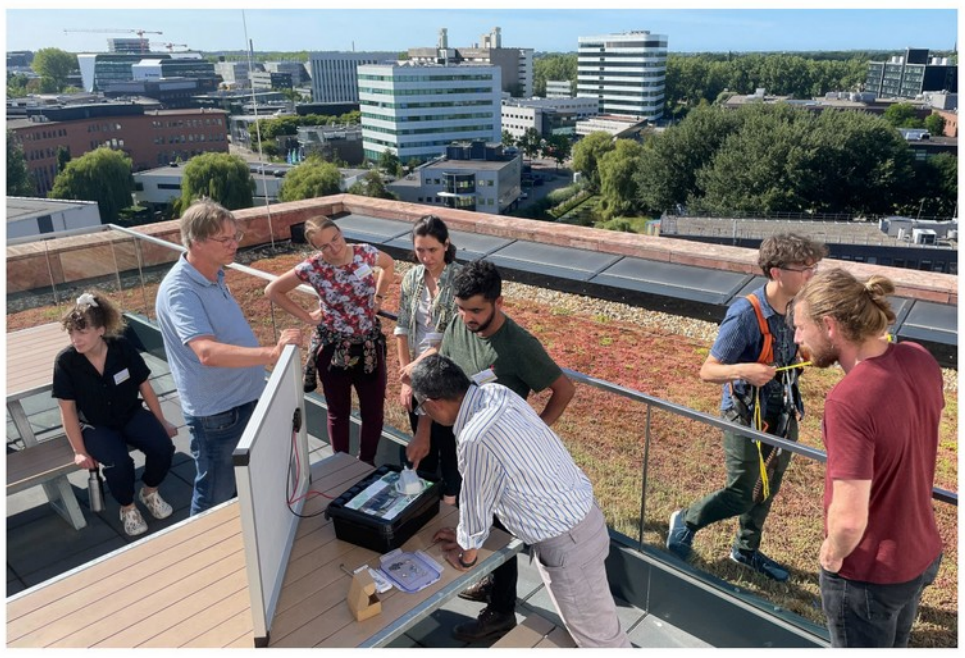
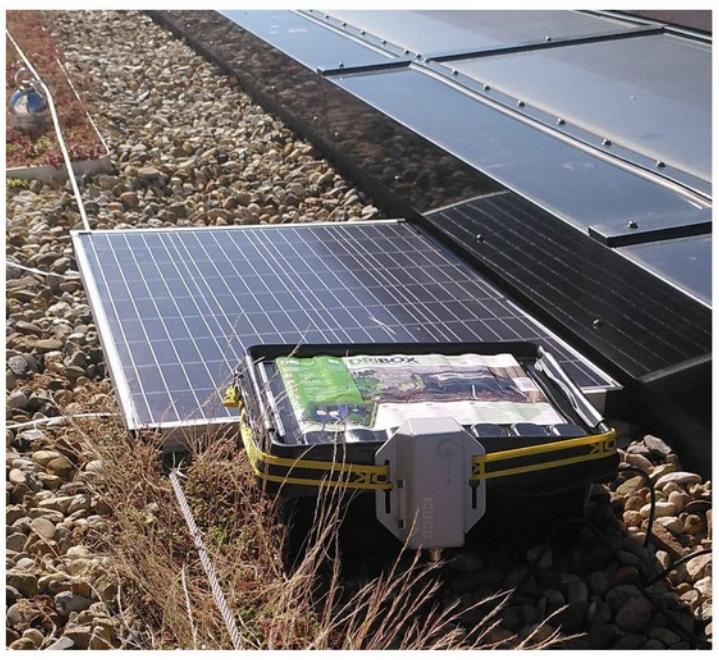
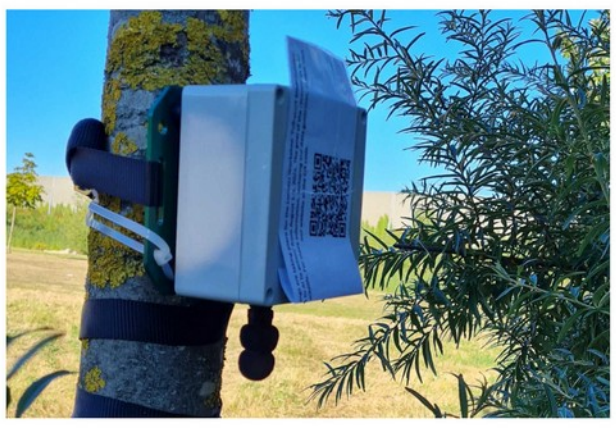
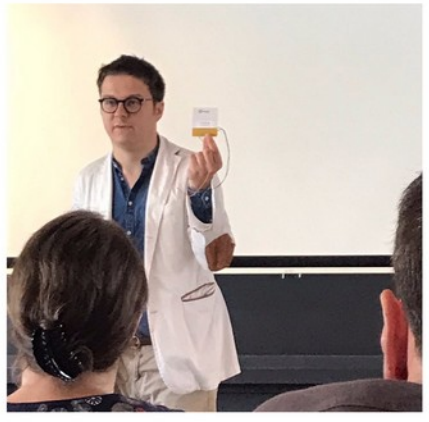
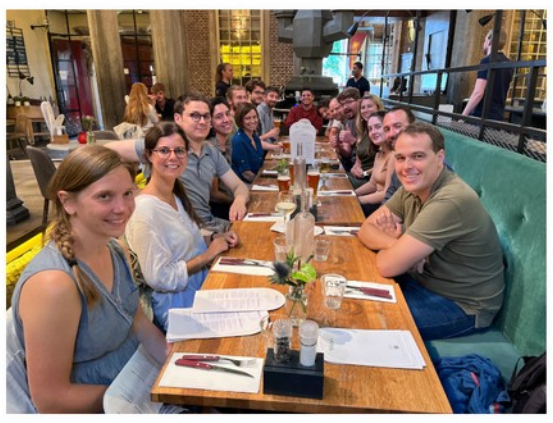
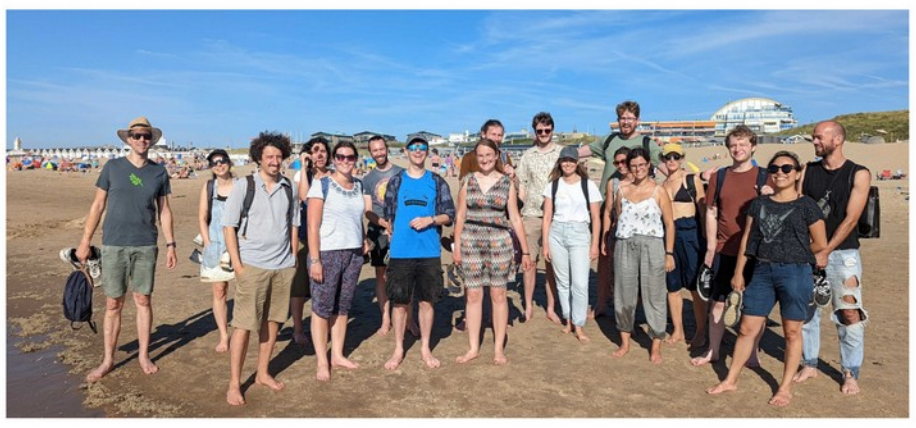

# Hardware platforms introduced/trialled

### Audiomoth - Peter Prince (Open Acoustic Devices)

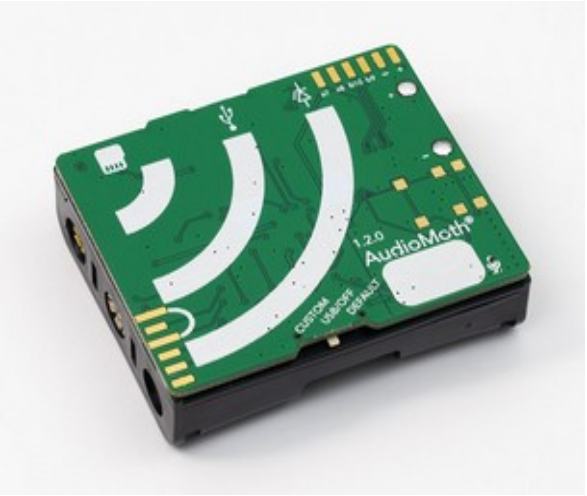

AudioMoth is a low-cost, open-source acoustic logger, capable of recording uncompressed audio at up to 384 kHz. Capable of scheduled recording and reacting to amplitude or frequency thresholds.

### BARD - Vincent Lostanlen

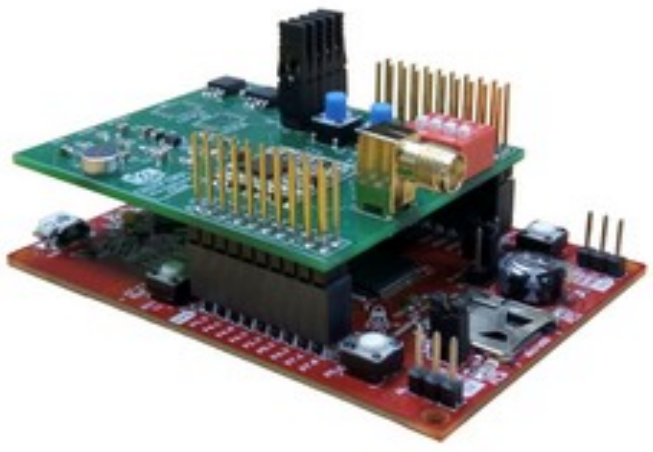

The BARD, or Batteryless Acoustic Recognition Device, aims to improve the autonomy and environmental footprint of passive acoustic monitoring. It comprises a microphone unit; a solar-powered microcontroller for real-time spectrogram analysis; a LoRa emitter for wireless communication; and non-volatile memory to recover from intermittent power losses.

### BEEP base - Pim Van Gennip

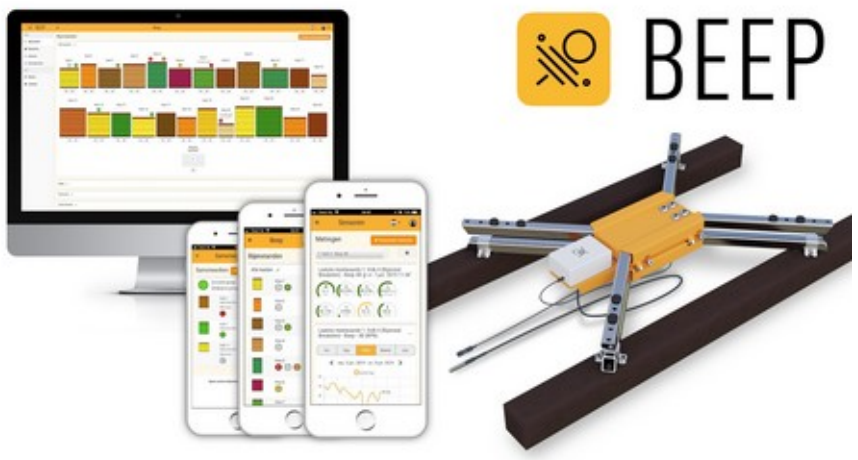

Website: https://beep.nl/home-english

Description: The BEEP base is an completely open source ultra-low power continuous sensor measurement device that measures the weight, temperature and sound spectrum of your bee hive. The built-in clock enables the system to turn on every 15 minutes to measure the values and to send the information to the free to use BEEP app through LoRa. All device settings are configurable via BLE and the iOS/Android BEEP base app.

### Healthy Climate Monitor - Pim Van Gennip

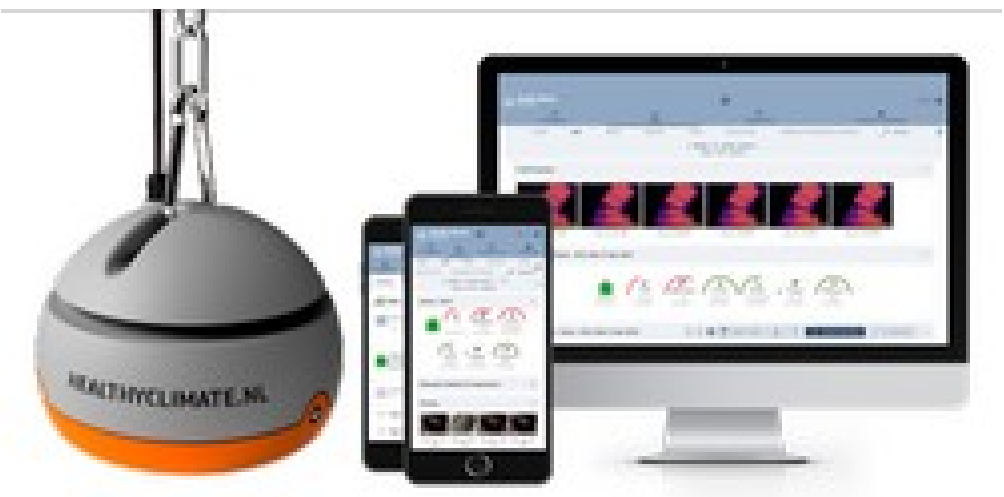

Website: https://healthyclimate.nl/home

Description: The Healthy Climate Monitor is a continuous sensor measurement device that is made for the harsh conditions of an indoor farm climate. It measures: temperature, relative humidity, carbon dioxide, ammonia, dust, air pressure, light, and movement and makes HD images day and night, has an optional FLIR camera and 48kHz 16 bit electret microphone. It sends its data every minute through 4G, WiFi, or ethernet connection to the Healthy Climate Monitor app.

## STREAMBOX - Grant Smith

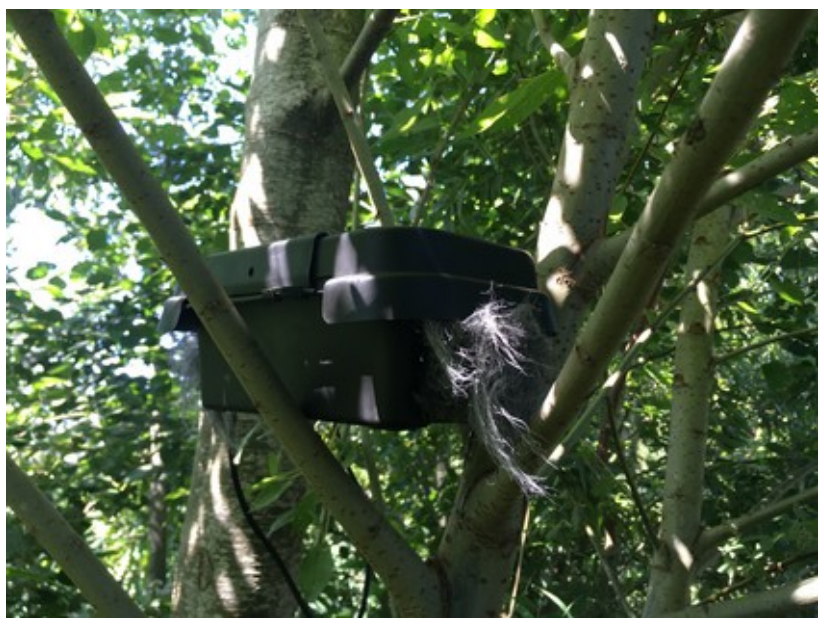

A streambox is a small weatherproof device that sends a live audio stream to a remote server, where it can be listened to on a public soundmap. The box can run on mains power, POE or a battery, which can be solar charged. It can connect via ethernet, wifi or a mobile network. A web interface gives access to settings such as gain, stream quality, mono / stereo and the exact location the box appears on the map.

Manual: https://paper.dropbox.com/doc/Streambox-with-IQ-Audio-card-and-Pi-Zero--BnLrv5QgktNA0~jqbmV7A0BdAQ-iWY7z0MaJYPCU6vfNs5UL
Soundmaphttps://www.locusonus.org/soundmap/

A project by Soundcamp with Locus Sonus and the Acoustic Commons network
soundtent.orglocusonus.orgacousticcommons.net

## Bugg - Sarab Sethi

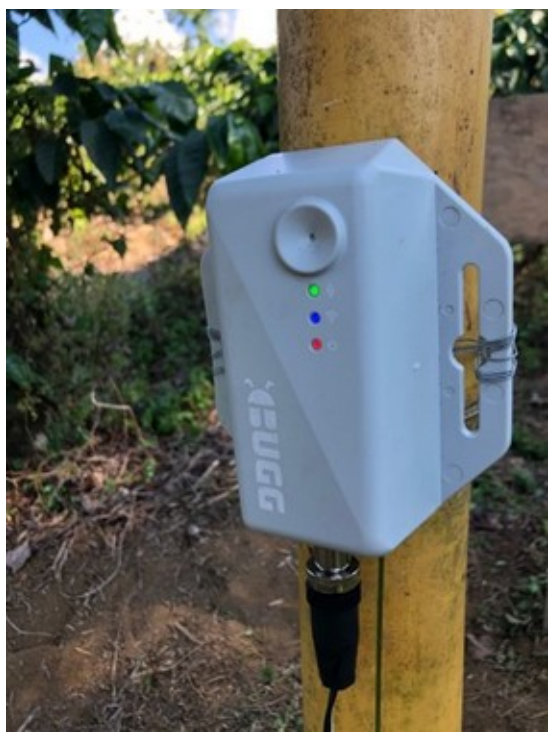

Website: www.bugg.xyz

Bugg is a device made for real-time long-term eco-acoustic monitoring. The device uploads data from the field over a mobile connection, and audio is analysed in the cloud and presented through a website

# Group reports

# Group-work overview: Group 1

**Ester Vidana-Vila** (La Salle Campus Barcelona, Universitat Ramon Llull); **Fredie Poznansky** (University of Exeter); **Grant Smith** (Soundcamp CIC); **Helen Whitehead** (University of Salford); **Joan Navarro** (La Salle, Universitat Ramon Llull); **Julian Zimmermann** (Max Planck Institute of Animal Behavior); **Lucia Manzanares** (Intituut Natuur en Bos); **Pavel Linhart (**University of South Bohemia)

# Motivation/aims

Group 1, as many other groups, was composed of a diverse group of people, with different backgrounds of expertise. As a result we decided to share our knowledge, learn new techniques and find commonalities in our work. As our main goal was to learn, we decided to work on insects because 1) we spotted orthoptera species at the location of recorder deployment and 2) to work with an animal group none of us was familiar with.

In this workshop we decided to check for the presence of crickets, bush crickets and grasshoppers in the soundscapes and to replicate classical observation that frequency of orthopteran sounds correlates with environmental temperature, while checking for bird, bats and human sounds. We used diverse expertise of each of the team members to do basic exploration of soundscape features (umap, ecoacoustic indices, false colour spectrograms) and sounds present in the recordings (manual annotations, BirdNet detections).

# Hardware and deployment details

We deployed three different devices in a water edge in Landskroon Park in Leiden from Monday until Wednesday (1-3 AUG) 2022:

**SwiftOne** (continuous recording in 10 minute file chunks, 48KHz sample rate, 28dB gain, https://www.birds.cornell.edu/ccb/swift-one/ ) - aimed to record terrestrial sounds and animals within human hearing range.

**Audiomoth** (1hr before sunset - 1h after sunrise, recording onset triggered by occurance of ultrasound frequencies and then records 60s, 192 kHz sample rate - good for all bat species in the netherlands, medium gain, highpass filter at 10kHz, https://www.openacousticdevices.info/audiomoth) - aimed at recording ultrasound bat sounds, e.g. Pipistrellus / Myotis bat species.

**Streambox** (live audio at 44.1kHz at around 320kbps, stereo with 2 x PUI capsules in a rough binaural arrangement, streaming via 4G with a USB modem, configuration via web interface over wifi network distributed by the modem).

# Analysis methods used

**More detailed description of workflow for each task can be found at:**
https://www.notion.so/Group-1-125bbe488a4348ffa5efd4034c963346

**Insect species survey** (Grant, others)**.** Lack of automatic tools. Photo-identification of species on site with iNaturalist / Google Lens. Comparing sounds with those found in publicly available web pages touching identification of orthopteran species (Recordings of UK crickets, bush crickets and grasshoppers - https://orthoptera.org.uk/node/707; Fraser's birding website lists characteristics of some orthopteran species http://www.fssbirding.org.uk/greatgreenbushcricketsonogram.htm).

**Bird species survey** (Helen, Pavel)**.** Manual annotation - Screening through the first 10 minutes chunk in each hour in RavenPro and annotating calling and singing bouts for each species. Unknown sounds pasted in web BirdNet tool and cross checked suggested species with examples on Xeno-Canto and species occurrence in eBird; Automatic annotation - Using BirdNet for screening all recordings. Comparison of annotations - Excel to compare species detected by 2 human observers, and BirdNet in annotated files; compare number of species detected in each recording chunk and their activity (duration of singing and calling bouts).

**Bat species survey** (Lucia)**.** Recorded files were checked manually in Kaleidoscope (couldn't download annotation as it was the free version). Automatic detection is possible in SonoChiro, Anabat (not recommended), Kaleidoscope Pro and Tadarida (you need to create your own dataset to identify species).

**Exploring soundscapes using acoustic indices** (Fredie, Ester, Joan)**.** 60 different acoustic indices were calculated for each 10 minute audio file using the Python package scikit-maad.

Ulloa, J. S., et al. (2021). scikit-maad: An open-source and modular toolbox for quantitative soundscape analysis in Python. Methods in Ecology and Evolution, 2041-210X.13711. https://doi.org/10.1111/2041-210X.13711

**Exploring soundscapes using false-colour spectrograms** (Joan, Ester, Fredie)**.** Basic false colour spectrogram was produced with the Python code by Sarab Sethi and Dan Stowell. https://github.com/sarabsethi/false_colour_index_spectrogram

**Exploring soundscapes using UMAP** (Julian). Segmentation of data into 2s clips. Mel Spectrogram for each clip. Calculate and visualise UMAPs with the method and Python scripts published in Thomas et al 2022. We checked examples from prominent clusters to establish what the cluster represents.

Thomas, M., et al. (2022). A practical guide for generating unsupervised, spectrogram-based latent space representations of animal vocalisations. Journal of Animal Ecology, 91(8), 1567–1581. https://doi.org/10.1111/1365-2656.13754

# What we learned

Setup different recorders for field deployment - recorder does not get stolen even if visible to by passers (might differ depending on country though…).

Sound characteristics of different animals - important for recording and analysis setup.

To visualise and evaluate dominant acoustic features of soundscapes (UMAP).

Visualise large ecoacoustic datasets (false colour spectrograms, ecoacoustic indices) to identify patterns in soundscapes.

To use BirdNet to screen avian species in long term recordings.

Impact of different sounds on soundscapes - low biophonic activity is hard to be detected by acoustic indices; bush cricket sounds and anthropogenic noise were the most prominent features captured by UMAP in our soundscapes.

Scale down expected outputs to match them with time available for the task.

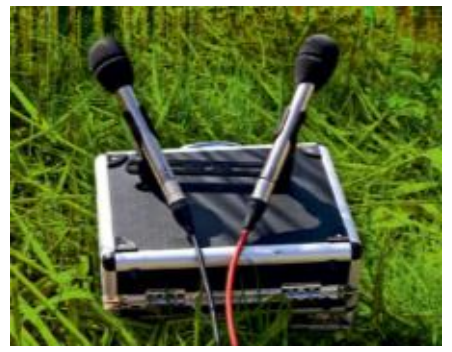

# *Using different hardware and analytic approaches to characterize the soundscape of Landskroon Park, Leiden*

## Full-stack bioacoustics: Field kit to AI to action workshop

| *Topics* | Device and location selection |
|---|---|
| **Cricket activity**<br><br>*Hanna told us that crickets change frequency with changing temperature!* | 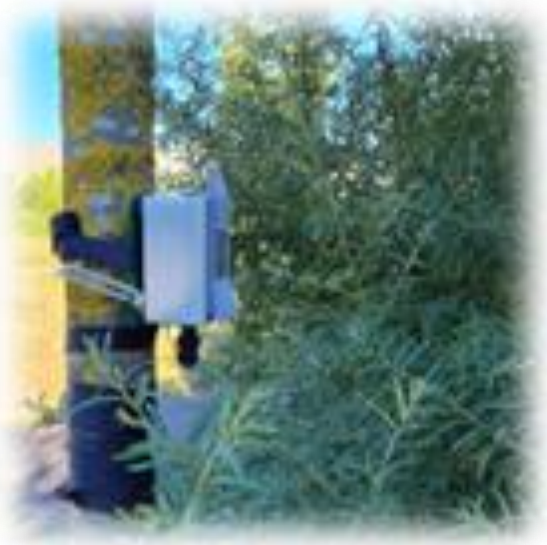 **Swift** |
| **Soundscape analysis**<br><br>*Which indices can be used to characterize sound analysis?* | |
| **Dimensionality reduction**<br><br>*Can we use dimensionality reduction to find interesting events?* | 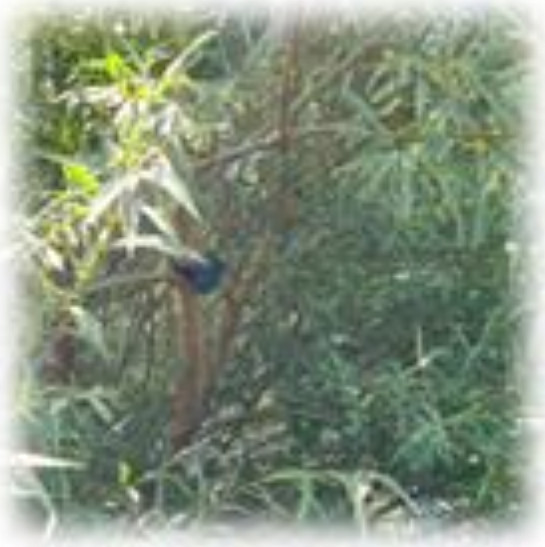 **Audiomoth** |
| **Bat presence**<br><br>*Are there any bats around?* | |
| **Livestreaming**<br><br>*Can we use livestreaming to explore a site remotely?* | 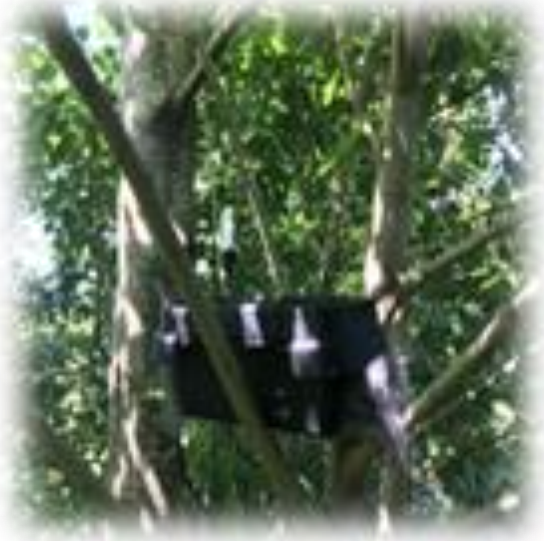 **Streambox** |

*Cricket activity*

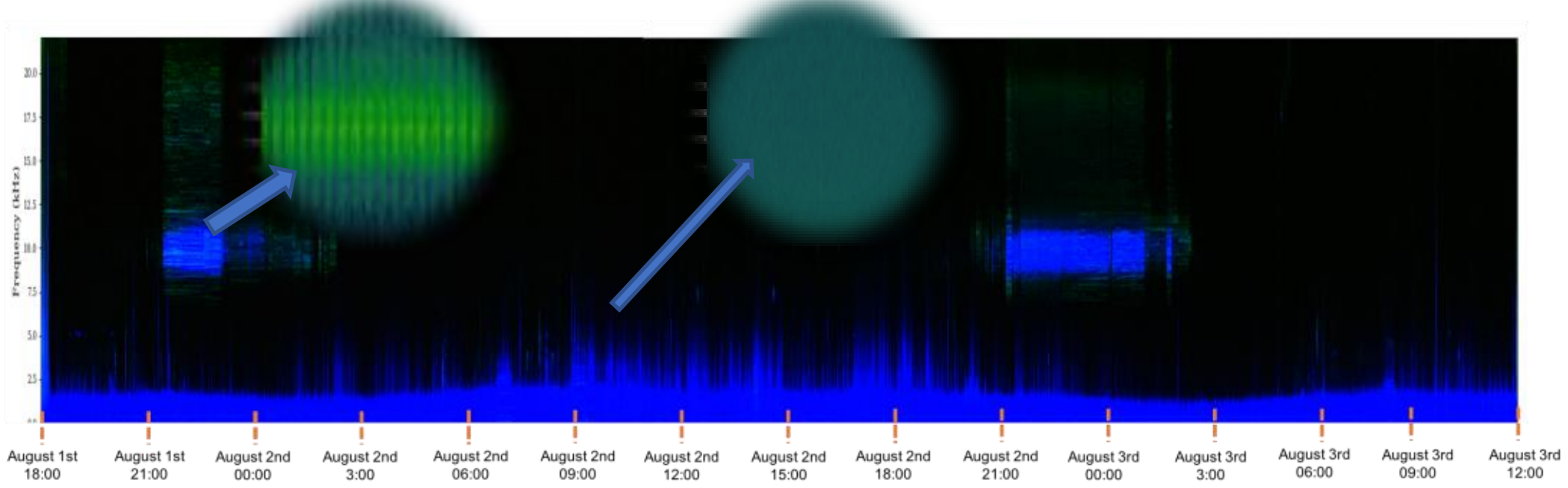

*Is UMAP able to maintain the human perceived similarity in the spectrograms?*

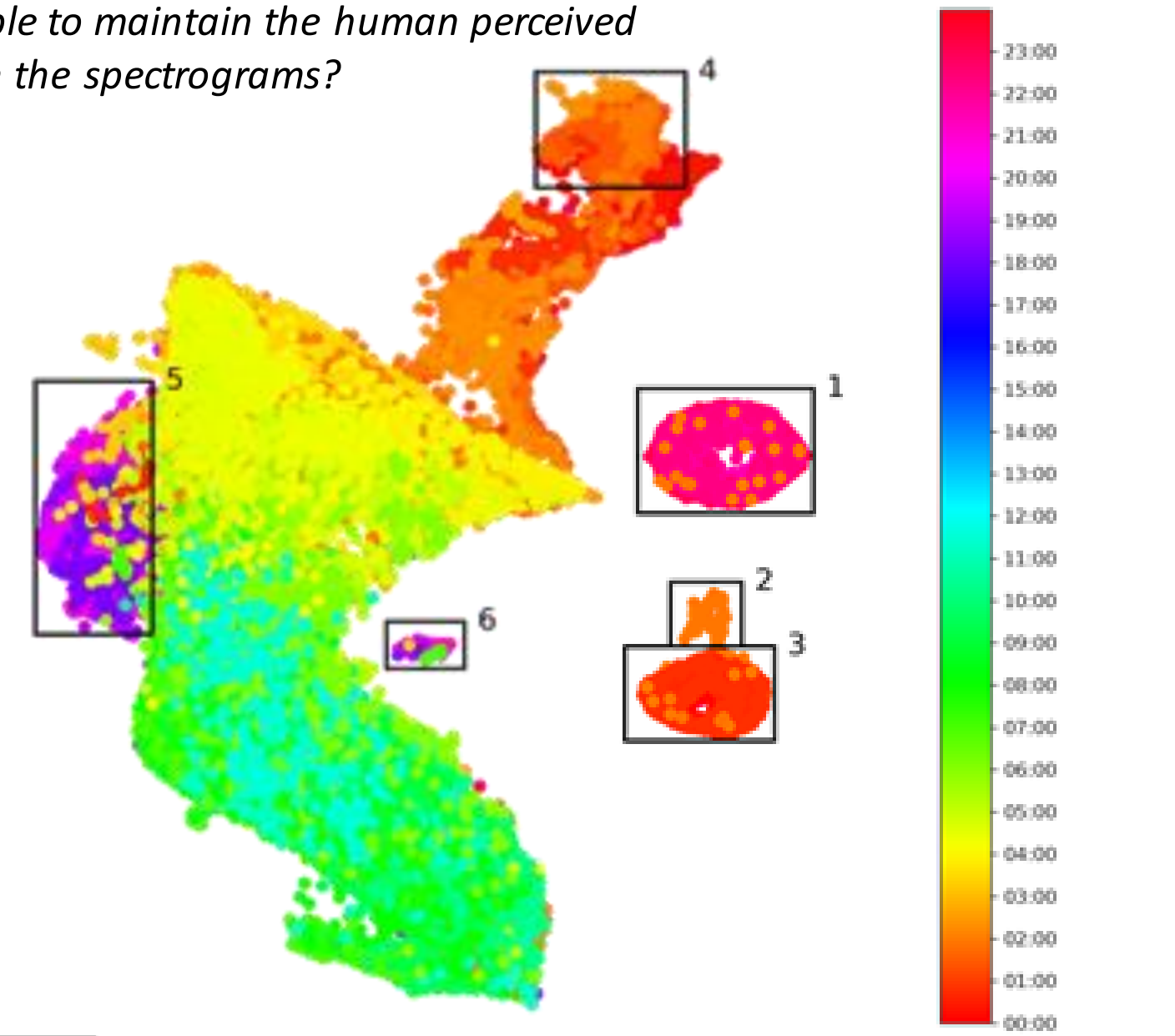

*Bird species count, 10 minutes annotation per hour.*

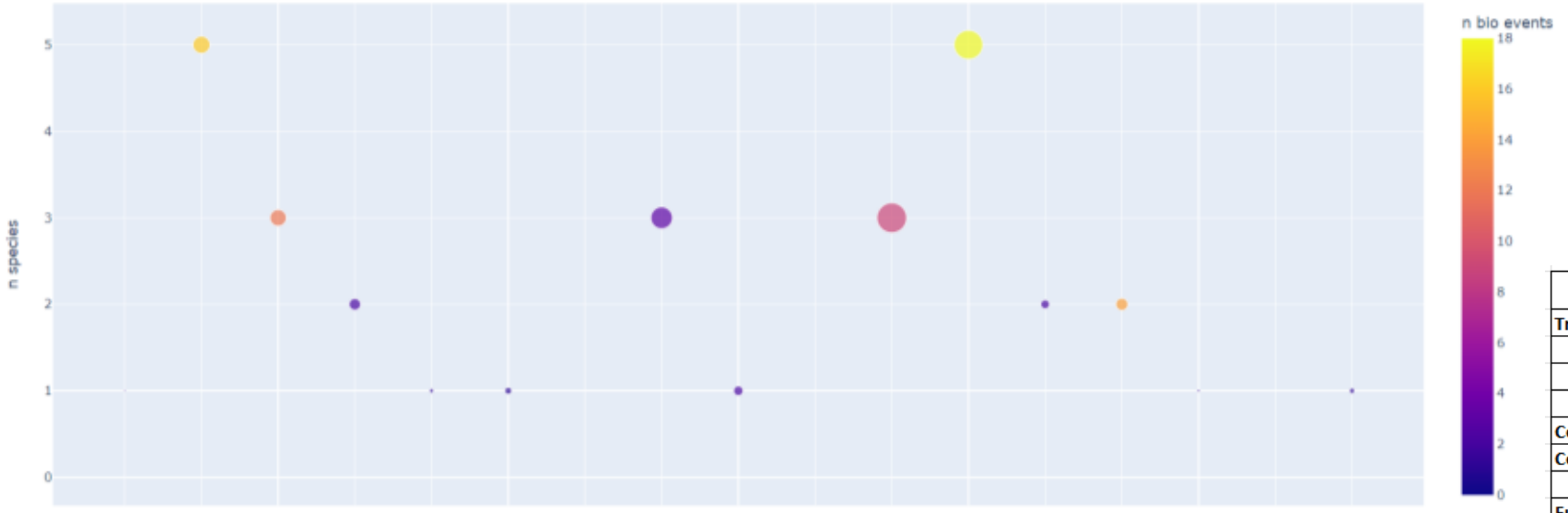

| Species list | | |
|---|---|---|
| **Truth - expert** | **Sloppy human** | **Birdnet** |
| | | Branta canadensis |
| | | Ardea cinerea |
| | | Chroicocephalus ridibundus |
| **Coleolus monedula** | Coleolus monedula | Coleolus monedula |
| **Corvus corone** | Corvus corone | Corvus corone |
| | | Corvus frugilegus |
| **Fulica atra** | Fulica atra | Fulica atra |
| **gull sp** | gull sp | |
| **Mareca strepera** | | Mareca strepera |
| | | Phasianus colchicus |
| **Phyllsocopus collybita** | Phyllsocopus collybita | Phyllsocopus collybita |
| **Psittacula krameri** | Psitaccus krameri | Psitaccus krameri |
| **Turdus merula** | | |
| **Vanellus vanellus** | | Vanellus vanellus |

| | |
|---|---|
| **true positive** | |
| **false positive** | |

*Do we have labeling bias?*

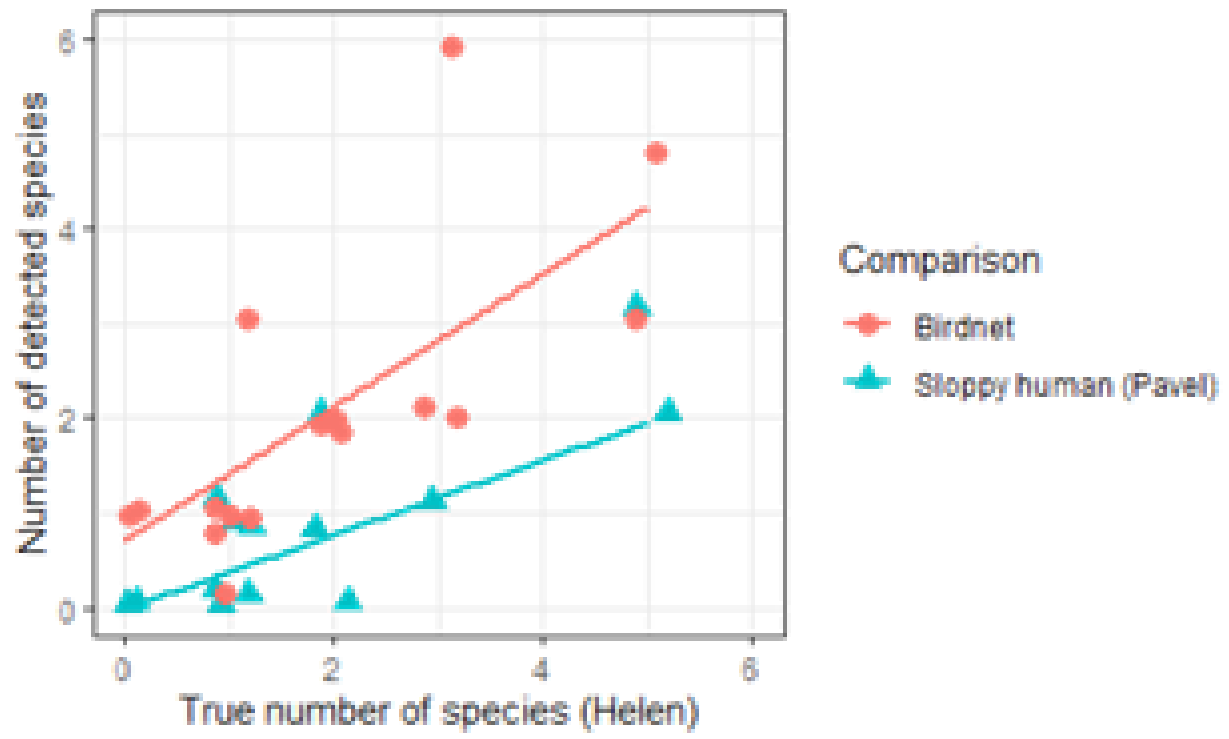

*Bioacoustic Index*

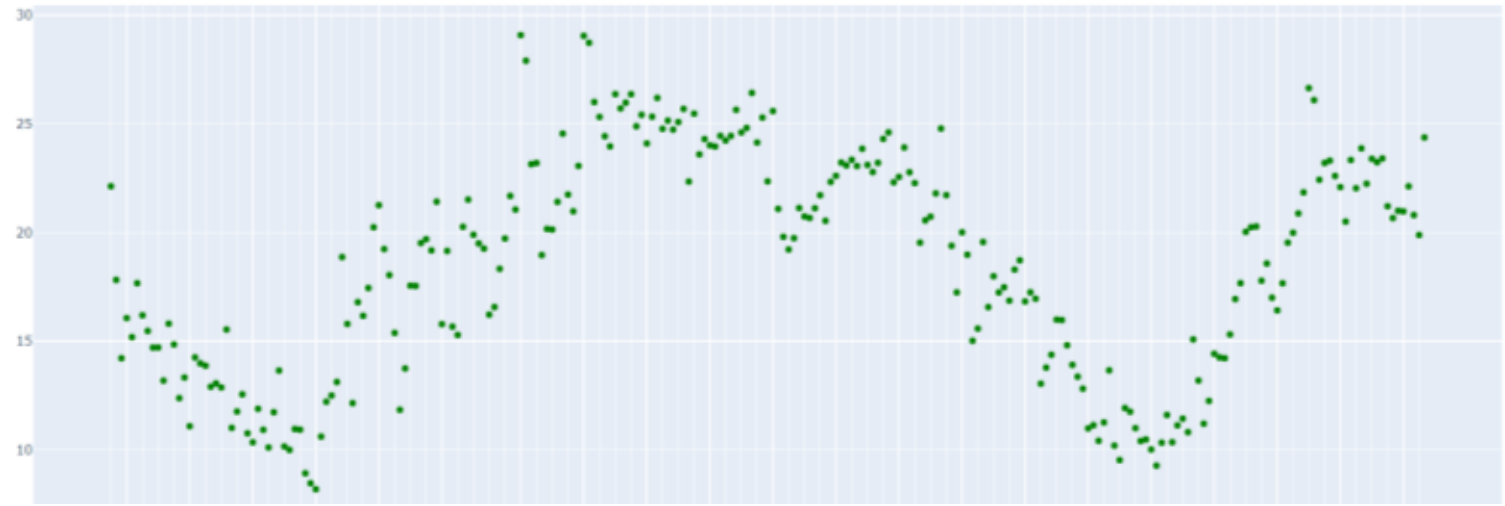

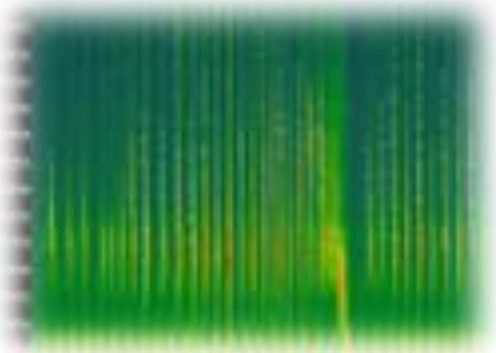

*Data processing obtained from Recordings Aug-1 to Aug-3*

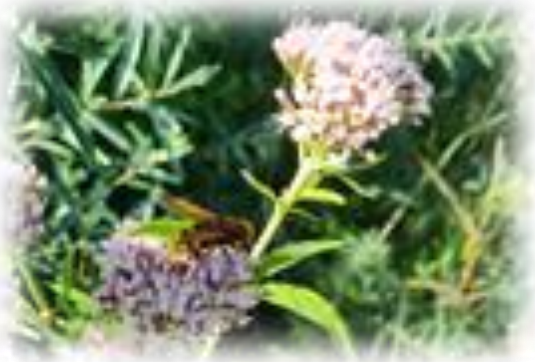

# Insights/hypothesis

- *Crickets change their vocalization frequency depending on the temperature.*
- *Acoustic indices allow us to characterize the soundscape.*
- *We have compared manual rapid annotation vs. BirdNet.*
- *We have used UMAP as a tool to analyze the data without listening to all the recordings.*

# Report (2pages)

**Group-work overview: Group 2**

Rogier Brussee

Ella Browning

Tom Bradfer-Lawrence

Hanna Pamuła

Ines Nolasco

Pim van Gennip

## Motivation/aims

Can we convey the experience of the Naturalis roof top through the analysis of its soundscape? This is aimed at a broad audience, both for researchers and as a science communication tool. The goal was to create an integrated experience, with temporally-explicit representation of the sounds on the Naturalis roof, shown through a variety of methods including acoustic indices, false-colour spectrograms, and the CityNet and BirdNet machine learning algorithms to identify the different sound sources in the recording.

## Hardware and deployment details

We used two recorders; an AudioMoth, powered with AA batteries, and a Bugg powered with a solar panel. We used sample rates of 48 and 44.1 kHz respectively. The AudioMoth recorded continuously for approximately 12 hours, and the Bugg recorded 5-minute files for 23 hours. The AudioMoth data were saved as a WAV file, whereas the Bugg produced mp3 files.

To create comparable recordings, we used *SOX* to split the single, ~12-hour AudioMoth WAV file into 5-minute WAV files, and converted the 5-minute Bugg mp3s to WAV format.

Slight differences in recorder positioning, and differences in device usage history led to very different soundscapes. The AudioMoth recordings were dominated by sound from the air conditioning.

## Analysis methods used

- Acoustic Indices: we used a custom function in R to calculate a suite of 11 acoustic indices (ACI, ADI, AEve, Bio, Bio sd, H, Hs, Ht, NDSI, NDSI.anthro, NDSI.bio). We generated time-series plots to (a) visually scan for outlier recordings which might have features of interest, (b) examine potential temporal trends, and (c) summarise soundscape patterns and sources of sound (air conditioning, birds and mammals etc). The differences in recording quality between the AudioMoth and Bugg was apparent in the indices values.
- False colour Spectrograms: we used code from Sarab Sethi to generate false-colour spectrograms, with Spectral Entropy, Acoustic Complexity Index and Spectral Power assigned to red, green and blue channels respectively.
- CityNet (Fairbrass et al 2018): CityNet consists of a pair of deep learning algorithms using convolutional neural nets. The algorithms were trained on annotated recordings collected in London UK. Types of sounds were annotated (bird song, insects, human speech, cars, planes, machine noise)and grouped into biotic and anthropogenic sounds. CityBioNet predicts the level of biotic sound; CityAnthroNet predicts the level of anthropogenic sound.

  We used Python 3 to run the CityNet algorithms on the 12 hour AudioMoth data and 23 hour Bugg data.  The levels of biotic and anthropogenic sounds in the recording were predicted.
- BirdNet (Kahl, 2020)
  - ResNets, trained on 984 species North American and European species
  - We've used the BirdNet-GUI, set up the coordinates for Leiden, week 31 (so the checklist of 113 birds was preselected)

## What we learned

- Acoustic indices
- False colour spectrograms
- How to use BirdNet and CityNet models
- Impact of sensor choice on results.

presentation with full results here : https://docs.google.com/presentation/d/1zwpyug84PrNTMa46Zo-7HswM0ItrylS6t5yLY8G2Kuc/edit?usp=sharing

## Acoustic Indices

AudioMoth recordings

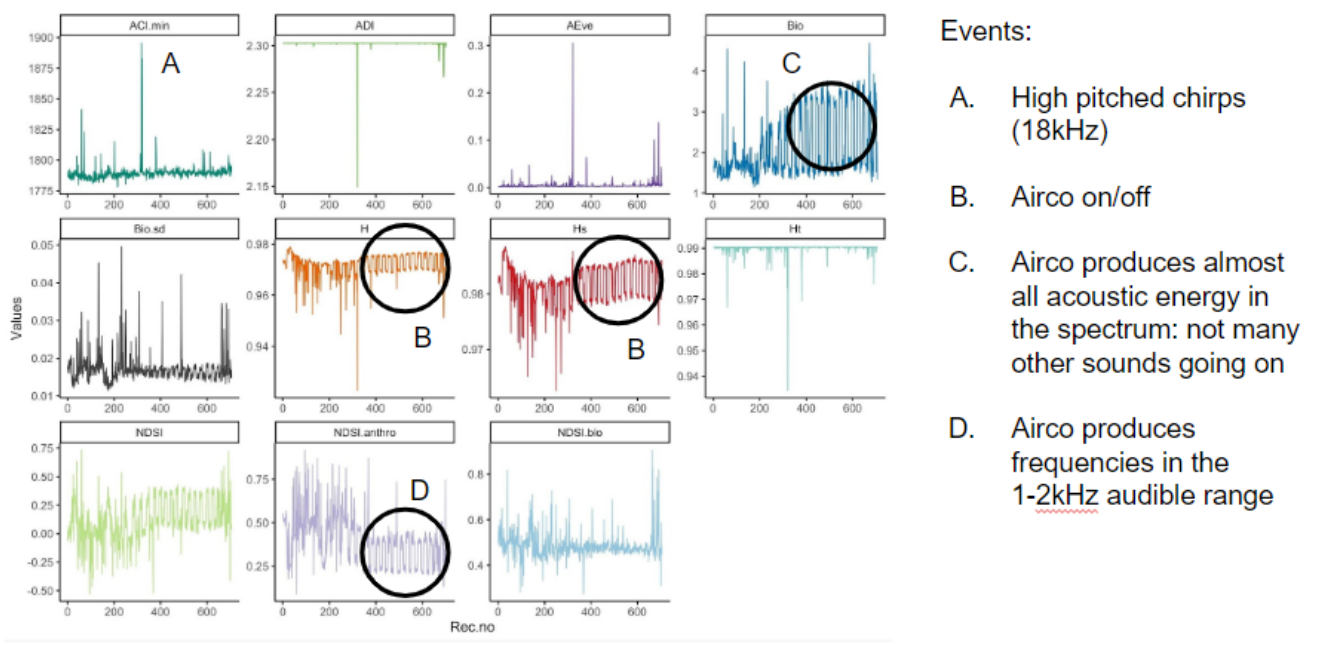

Bugg recordings:

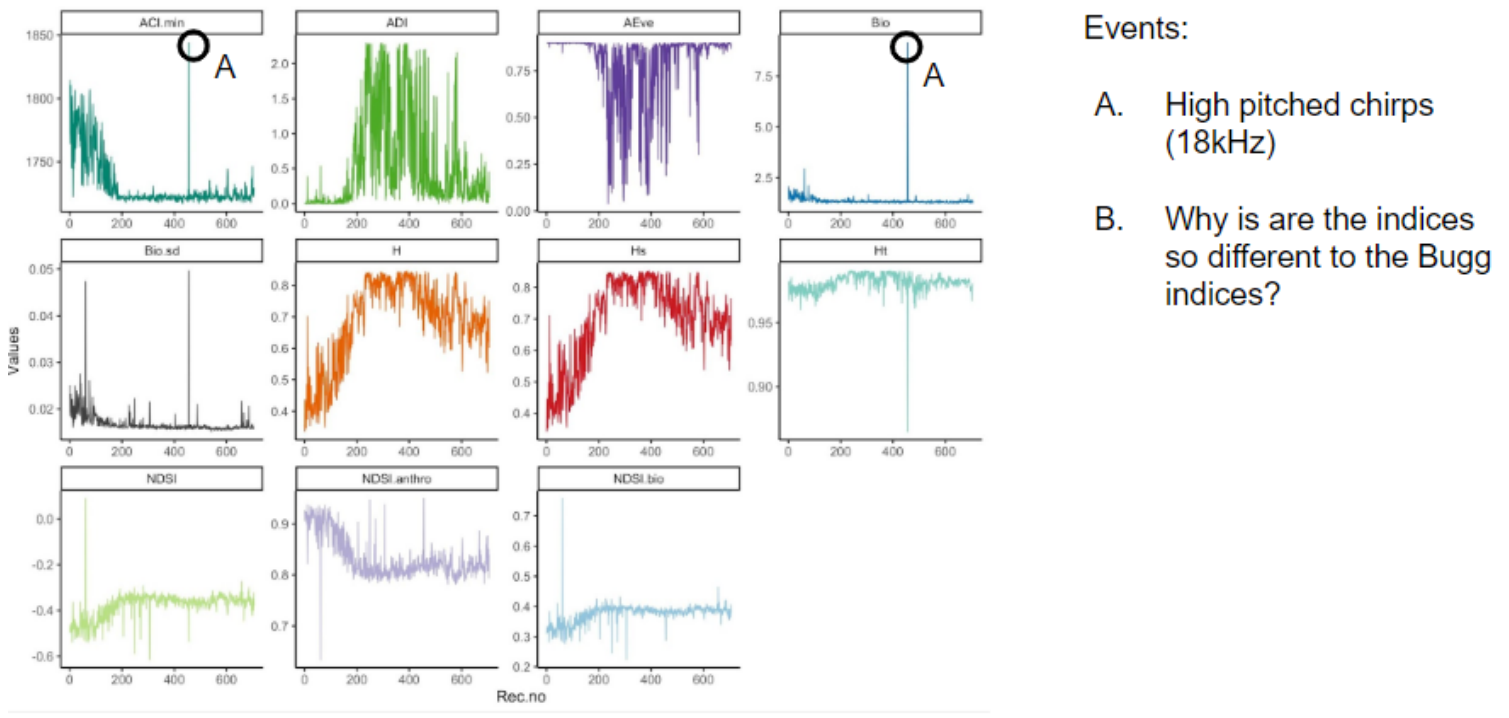

## False colour spectrograms

AudioMoth :

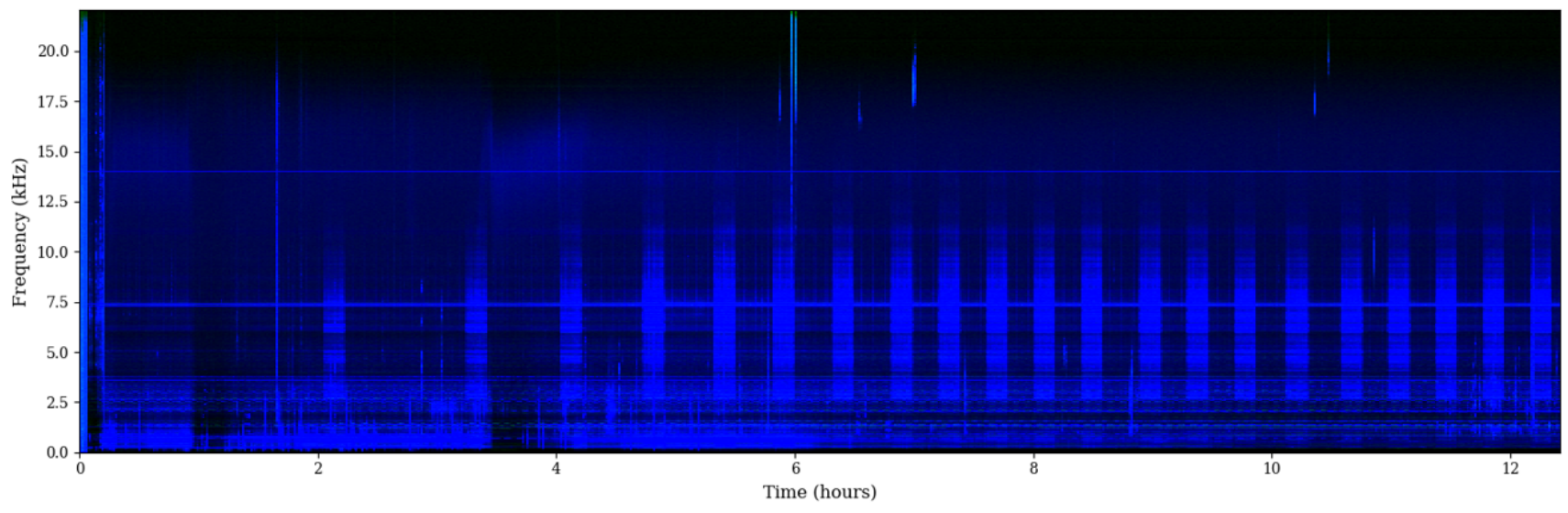

Bugg recordings:

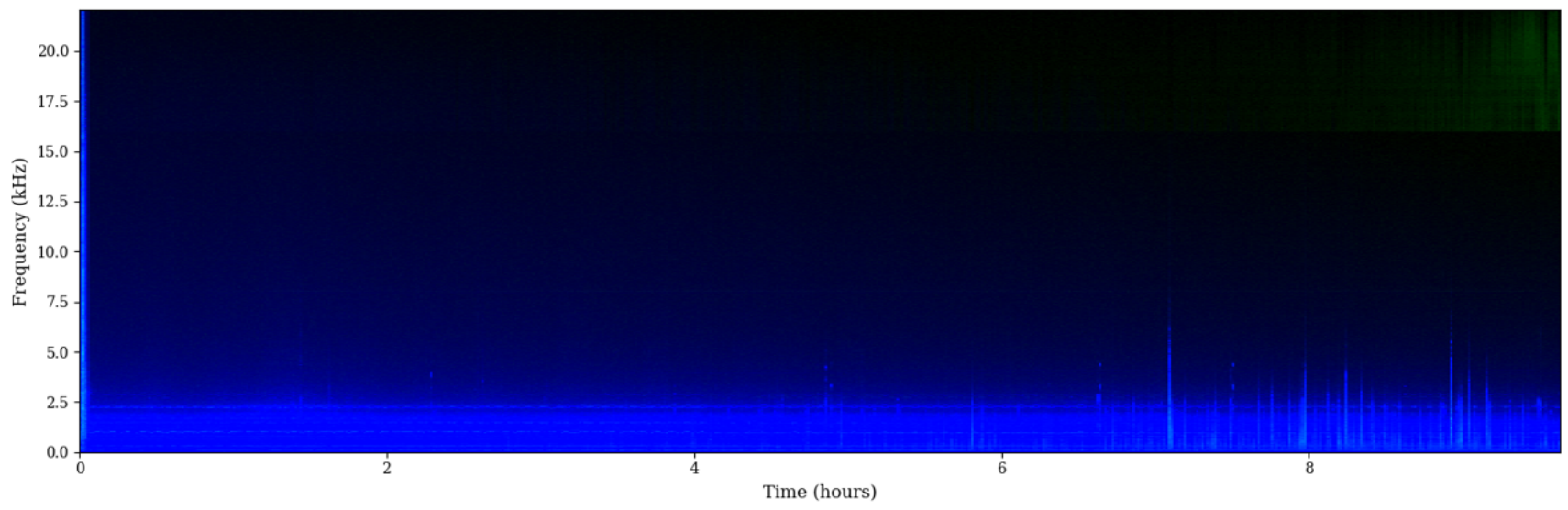

## BirdNet Results

number of bird detections over 24 hours:

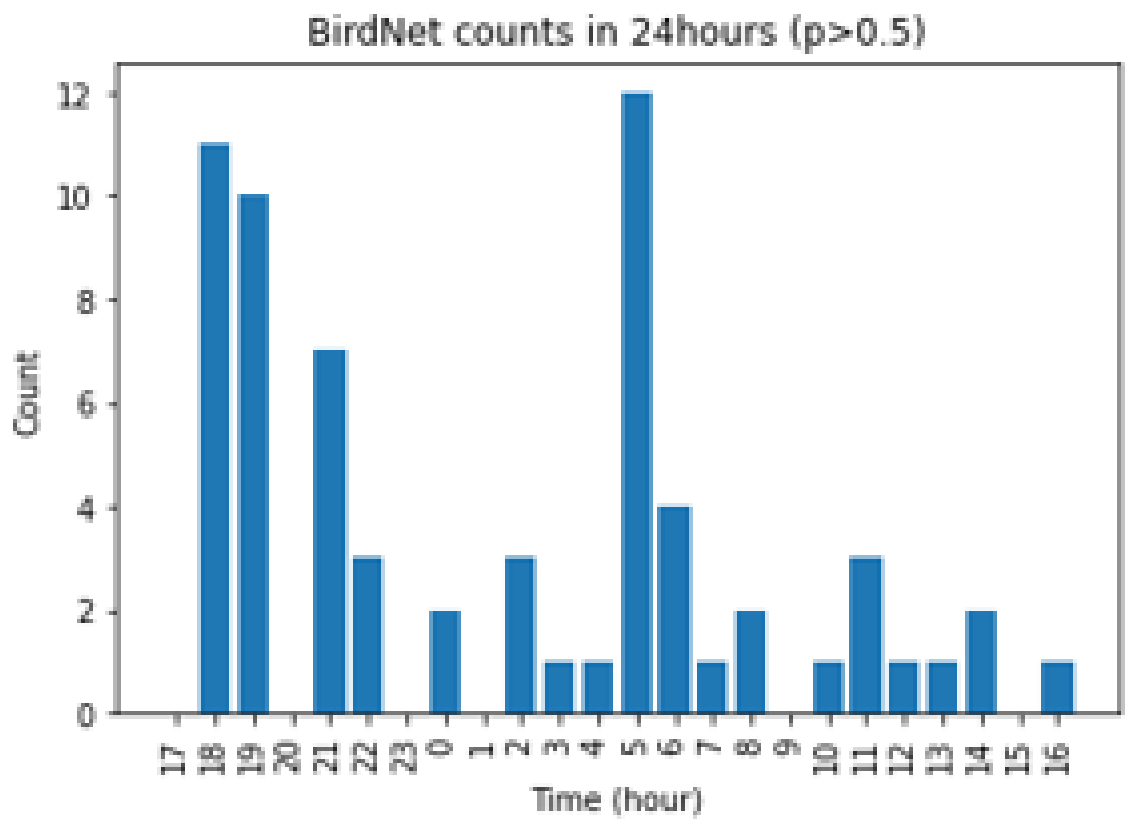

Some of the bird species detected on recordings from Bugg and AudioMoth devices (Bugg detections were manually validated!) :

Bugg:

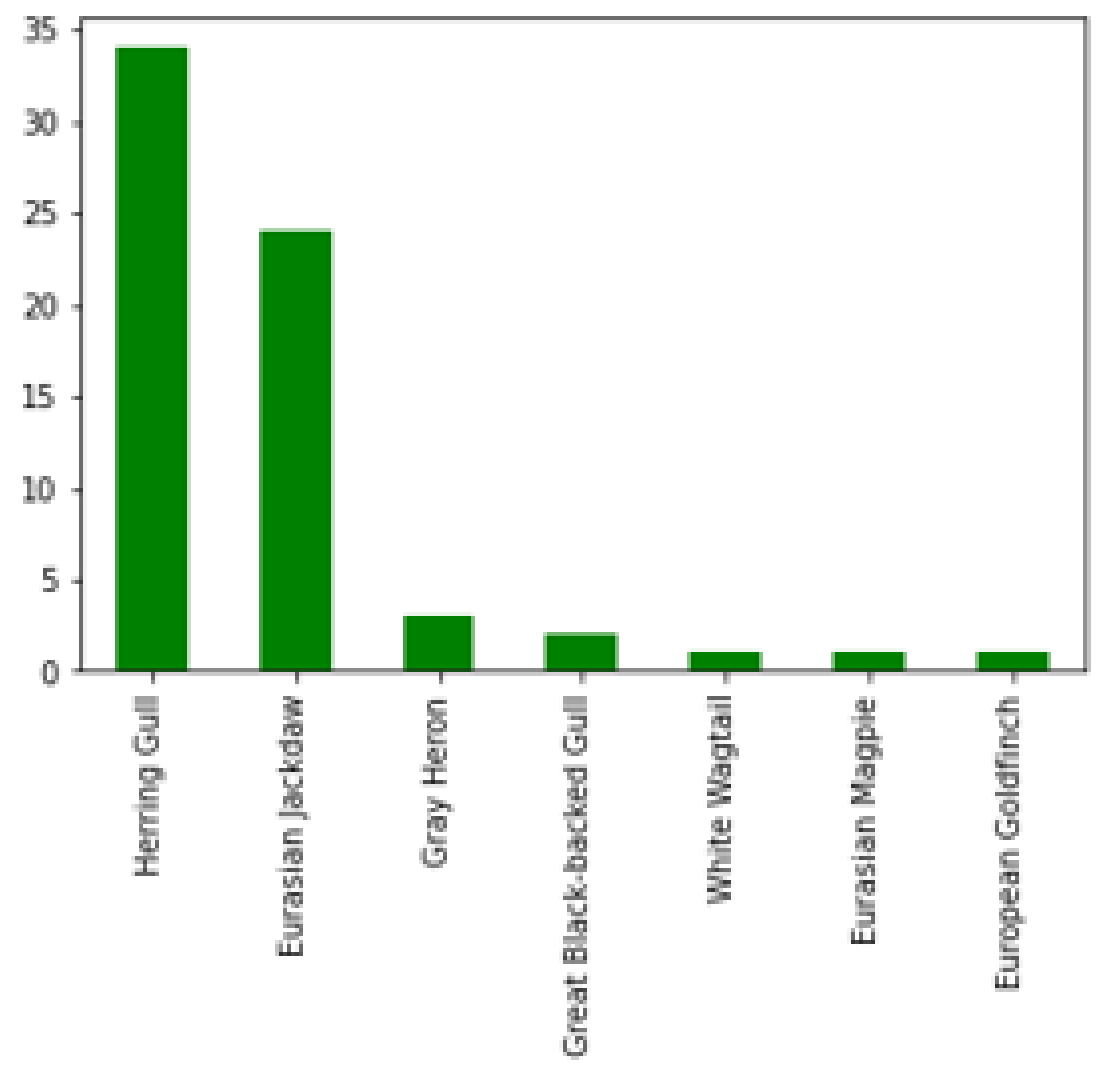

AudioMoth:

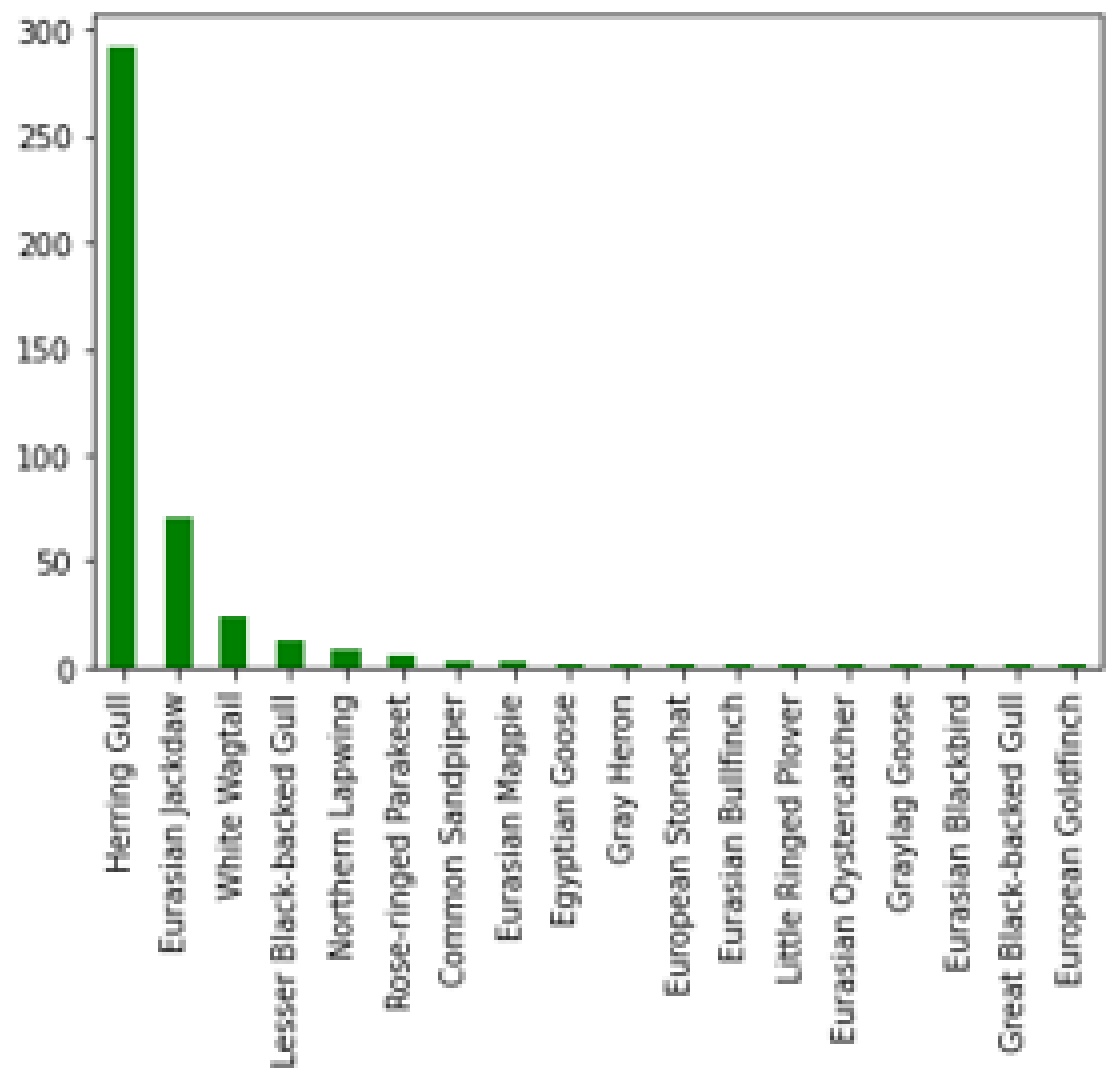

## CityNet results

CityBioNet predicted mean biotic sound levels in the AudioMoth data of 0.35 (sd 0.06), whereas the mean predicted biotic sound levels was 0.03 (sd 0.02) in the Bugg data. CityAnthroNet predicted near constant anthropogenic sound, (mean 0.99; sd 0.01) in the AudioMoth data. In contrast, anthropogenic sound levels were predicted to be low in the Bugg data (mean 0.06; sd 0.10), although there were peaks in anthropogenic sound levels  throughout recording period.

The significant differences in predicted levels of biotic and anthropogenic sounds levels between the AudioMoth and Bugg data could be due to a range of factors, including deployment position, differences in microphone sensitivities, device noise, or conversion of compressed Bugg mp3 files to WAVs.

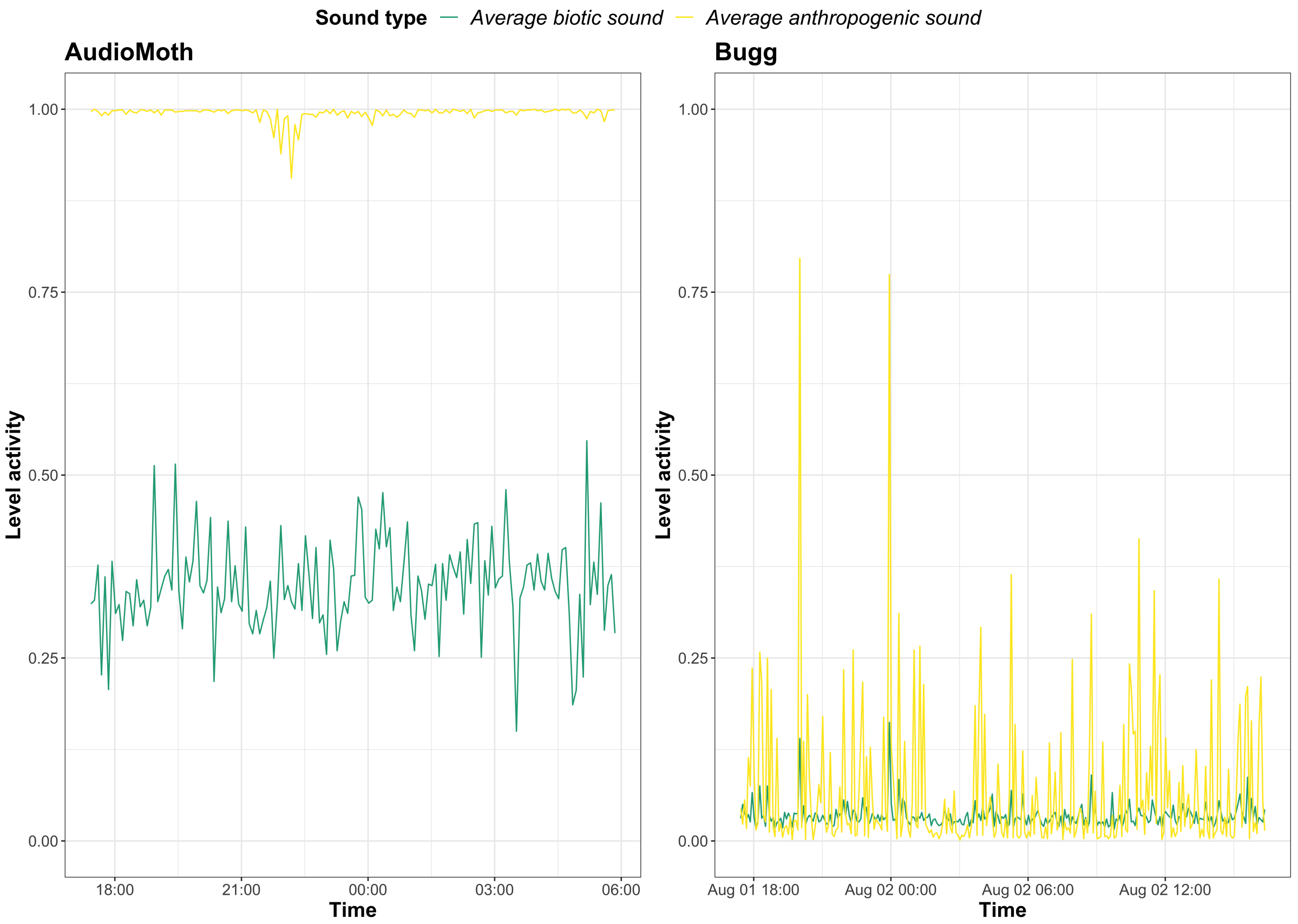

Figure 1: Predicted levels of biotic and anthropogenic acitivty in AudioMoth (right) and Bugg (left) recordings.

# Group-work overview: Group 3

**Members:** Benjamin Cretois, Felix Michaud, Ivano Pelicella, Padmanabhan Rajan, Sylvain Haupert, Tereza Petruskova

## Motivation/aims

The main objective was to analyse the soundscape on the Naturalis roof. We especially wanted to estimate a few metrics of biodiversity using sound (birds and/or insects.)

- Detection and discrimination of natual sound coming from animals and non biological source (wind, and human noise.)
- Species detection via BirdNet
- Acoustic indices computation and correlation in order to understand sound biodiversity.

A second objective was to test if the algorithm designed by Felix can automatically detect syllables of the yellowhammer (*Emberiza citrinella*) song, which are used for individual identification and subsequently for individual acoustic monitoring of this species. Data was provided by Tereza - directional recordings of 23 yellowhammer males recorded in 2019, each male was represented by one recording with a different number of songs. A variant of the DB-Scan algorithm was used for this.

## Hardware and deployment details (Naturalis data)

1 AudioMoth device, 1 BUGG device deployed on the roof of Naturalis Museum for 23 hours from August 6th 2022 till August 7th 2022.

Sampling rate 48 Khz single channel

The BUGG used compression (data saved as mp3 files)

Settings: continuous recording

File splitting 5 minutes

**ps:** After the data was retrieved, we found that the Audiomoth resulted in several empty wav files, and a single 4 GB wav file. Moreover, the BUGG was unable to transmit the collected data over the cellular network.

## Analysis methods used

For soundscape analysis, the following was performed:

- Manual annotations of the first minute of every quarter of hour
- Anonymization of the dataset using Voice Activity Detection algorithm (i.e. WebRTC VAD; https://gitlab.com/nina-data/eco-acoustic)
- Analysis of avian biodiversity using BirdNet [1] on the entire duration of the audio files collected over the 23 hours
- Computation of acoustic indices on the first minute of every quarter of hour (over the

23 hours) using scikit-maad Python package [2] following the tutorial found here

Please find some figures that summarise the data in the presentation:
https://docs.google.com/presentation/d/1B5tGkZEDvr19tpcjcKUk6w0QlfUwyp4GIA24Wp-74qY/edit?usp=sharing

- Manual annotations can be found here: labels
- The 1 minutes files that were used in the analysis can be found at this dropbox link: https://www.dropbox.com/s/7oaklgm2pkpbbnu/files_anonymized_1min.tar.gz?dl=0

# What we learned

- How to configure and deploy an autonomous recording device
- How to manually annotate the audio files (we used the first minute every 15 minutes). This was done with a free software (Audacity) to visualise spectrograms and to find out significative information identified with headphones
- Process the data
    - Elaboration of compressed audio file with BirdNet and species identification
    - Birdnet => get the number of birds per file to estimate part of the biophony
    - scikit-maad => get the indices and display them
- Human speech detection & anonymisation: A simple GMM-based speech detector was used: WebRTC VAD (two classes: speech/non-speech.)
  The assumption here was that the human speech was the loudest in the recordings.
- An energy-based segmentation method was also tried (30% of average short-time energy of the entire file was used as a threshold to discard non-speech.) This resulted in a number of false alarms.
- The mp3 files were converted into wav files and resampled to 16 kHz. This probably resulted in a loss of information. BirdNet detected “seagull” as “coyote”. Nevertheless, after using BirdNET on the original files that were collected during the field experiment, BirdNET’s classification returned coherent results.
- Most of the 58 acoustic indices are strongly correlated because most of the 1 min audio files were empty, containing only wind blows. Moreover, it is difficult to observe the dawn/dusk chorus from the indices probably because very few bird calls/songs were captured by the recorder (i.e, from manual annotations we found that only 8.7% of the 1 min audio files contains biophony, mostly herring gulls and eurasian jackdaws).
- Most of the acoustic indices were related to anthropophony (planes, cars…) and geophony (wind) with a constant increase of the anthropophony/geophony from early morning (around 7-8am) till mid-day before decreasing till late afternoon (around 6-7pm).
- The false color spectrogram computed with 3 indices (mean, variance and skewness) gives a quick insight into 23h of the Naturalis soundscape. It is possible to distinguish the stationary sounds (in green, mostly due to anthropophony) to the discrete sounds (in red, mostly wind blow, sometimes birds)

Overall, we felt that the experiment we performed was not well thought out. We were more focused on the data collection and analysis. We should have had a clear objective and then decided what data to collect, where and when, and choose the best set of tools (manual annotations, acoustic indices, species identification, sound types clustering…) depending on the initial objective.

**Summary about the second objective:**

Felix managed to run first analyses resulting in three clusters, however, after Tereza's check, it was clear that the algorithm did not assign different syllables correctly. But both believe that they managed to identify the problems, and they plan to continue in this task also after the end of the workshop.

**Additional material (beyond 2 pages)**

**Personal observations :**

- Padmanabhan Rajan : One interesting observation I had was that many ecologists and data scientists have taken well to deep learning. Though skepticisms remain on the use of 'black-box' AI, tools like BirdNet are used to support many hypotheses.
- Sylvain Haupert : I really appreciated working with people coming from very different backgrounds as everyone came with different objectives and different ways to process the data.
- Tereza Petruskova : Same as for Sylvan, I really enjoyed seeing the other perspectives on bioacoustic research. Moreover, working together with Felix has assured me again, how crucial is collaboration among biologist/ecologist and "IT people" as some goals cannot be achieved without "seeing both sides of the story".

# References

[1] Kahl, S., Wood, C. M., Eibl, M., & Klinck, H. (2021). BirdNET: A deep learning solution for avian diversity monitoring. *Ecological Informatics*, *61*, 101236.

[2] Ulloa, J. S., Haupert, S., Latorre, J. F., Aubin, T., & Sueur, J. (2021). scikit-maad: an open-source and modular toolbox for quantitative soundscape analysis in Python. *Methods in Ecology and Evolution*, *12*(12), 2334-2340.

# Group-work overview: Group 4

Lisa Gill, Claire Hermans, Kinga Jánossy, Vincent Lostanlen, Oliver Metcalf, Richa Singh

## Motivation/aims

**Comparison of Bugg vs AudioMoth audio data**

The aim of this project is to find a way to compare audio data from different hardwares. This would help to get comparable data throughout a data collection campaign where settings, devices, etc might be changed during the sampling period.

## Hardware and deployment details

We deployed two acoustic monitoring devices, a Bugg and a AudioMoth next to each other to get synchronized recordings. The Bugg device had 48 kHz sampling rate and files were compressed into MP3 files, while the AudioMoth recorded raw WAV files at a 44.1 kHz sampling rate.

We first recorded natural soundscape along a canal for two hours (12 to 2 pm), followed by 10 min playback of Golden Oriole and Great Reed Warbler (audios from Xeno-Canto). For the playback experiment, we also recorded via the DawnChorus app to add an extra hardware for comparison.

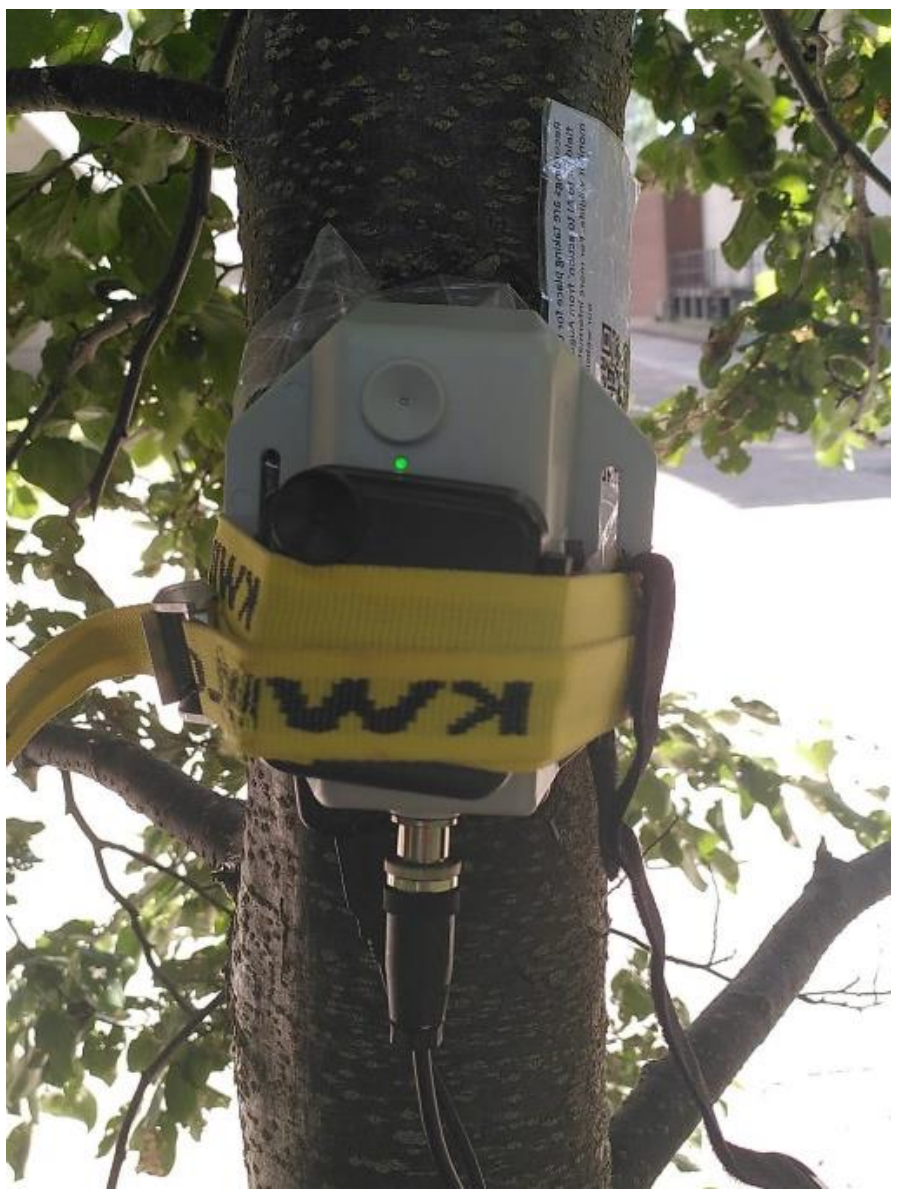

## Analysis methods used

Several steps were taken for comparison:

1) Bugg's MP3 data were converted to WAV files.
2) We aligned 5 min chunks of recordings from both devices.
3) We downsampled the Bugg's data to 44.1 kHz.
4) We applied a high-pass filter at 200 Hz and normalized the data.

Methods to compare the recordings:

1) Spectrograms
2) Power Spectral Density (PSD)
3) Acoustic indices on raw data and HP-filtered and normalized data:
- Entropy (H)
- Normalized Difference Soundscape Index (NDSI)
- Acoustic Complexity Index (ACI)
4) Signal-to-Noise Ratio (SNR)
5) BirdNet Analyzer for species identification on the playback recordings.

Extra exploratory analyses can be found on this repository:
https://github.com/kngjnssy/full_stack_bioacoustics

# What we learned

It is important to select appropriate hardware depending on the research questions and species/soundscape of interest and to have regular check of the device's settings.

Check a first batch of recordings before deploying the devices for a longer period.

We learned how to work with acoustic indices, BirdNet Analyzer, MAAD, Source separation using Bird MixIT.

Seasonal timing is important in BirdNet Analyzer. For instance, when we played Golden Oriole playbacks and configured the BirdNet date settings in August, the BirdNet Analyzer couldn't identify the species due to absence of call activity of Golden Oriole in August. What if climate change affects seasonal timing in bird songs for example?

Before source separation, BirdNet Analyzer missed or got low detection of background species. The detection got slightly better after source separation. However BirdNet could not distinguish between two species. There might be some sound alteration that modifies the bird calls. Source separation could be helpful at the family/genus level rather than the species level.

### Group 6

## Is anybeing there?: An interactive workflow for exploring mysterious soundscapes events using learned embeddings: VGGish vs CPC

Avery Bick, Camille Desjonqueres, Alice Eldridge, Becky Heath, Jacob Kamminga, Vincent Kather, Peter Prince, Hendrick Reers.

## Summary

- Freshwater soundscapes are poorly understood.
- Rapid advances in ML are appealing, but lack of knowledge and freshwater datasets limit the application of supervised approaches
- We created a pipeline to identify freshwater soundscape types using pre-trained feature extractors and an interactive UMAP
- The tool afforded rapid exploration of 24hour soundscape recordings which enabled identification of sparse nocturnal freshwater invertebrates
- Future work will i) add audio playback to interactive UMAP ii) evaluate value of different embeddings and transfer learning

### Motivation & Aims

Identifying indicators to monitor and assess the effects of anthropogenic stressors on the ecological status of freshwater ecosystems remains a challenge. Passive acoustic monitoring shows promise to detect invertebrate species (van der Lee et al 2020). Deep learning is advancing bioacoustics and shows ecoacoustics, however lack of understanding of freshwater invertebrate calls precludes the use of supervised learning.

**Aim.** Develop an end-to-end workflow to support interactive exploration and interpretation of soundscapes using pre-trained DL embeddings

#### Ecological questions

- What are the differences in diurnal patterns between soundscapes above and below water?
- To what degree do terrestrial anthropophonic events enter the aquatic soundscapes?

#### Technical and workflow questions

- Which pre-trained embeddings are most valuable soundscape descriptors?
    - VGGish (Audioset)
    - Modified CPC (Librilight)

### Hardware and deployment details

**Study site:** Small body of water on the perimeter of Leiden University adjoining a bike path and road (52.168-4.458).

**Devices**: Hydromoth (v1.2.0) (192kHz SR, 16 bit) attached to a stick 20 cm below the water surface, audiomoth (v1.2.0)(192kHz SR, 16 bit) attached 1m above water level tied to a tree.

**Recording schedule:** 10 mins every 15 min over 24 hours

**Raw Data**: 24hr cycles for each device from 12:00 on 02/08/2022 to 11:45 on 03/08/2022

- Sample recordings from the pond

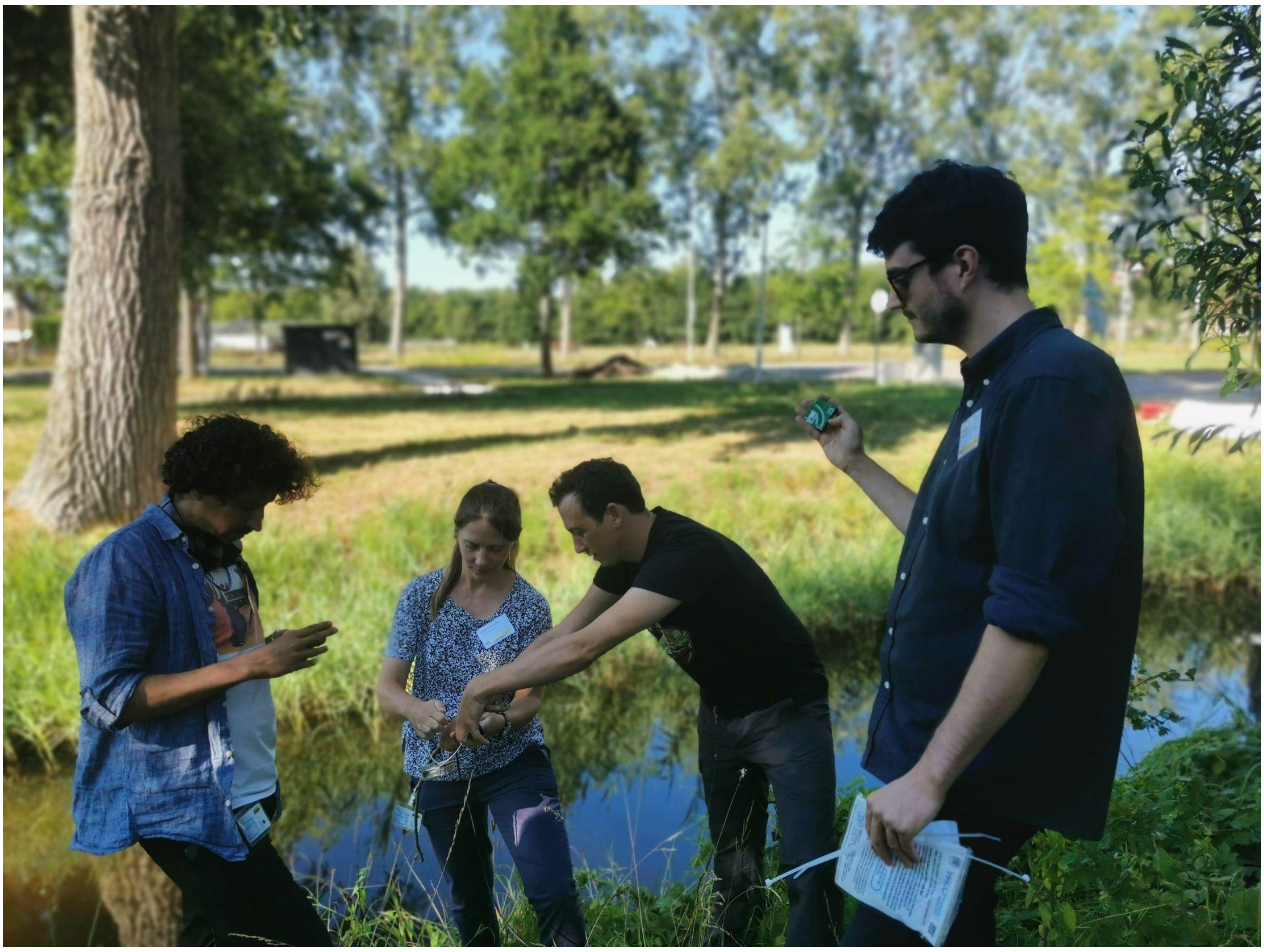

Fig. 1. Team members setting up audiomoth and hydromoth on Leiden University Campus

**Analysis pipeline**

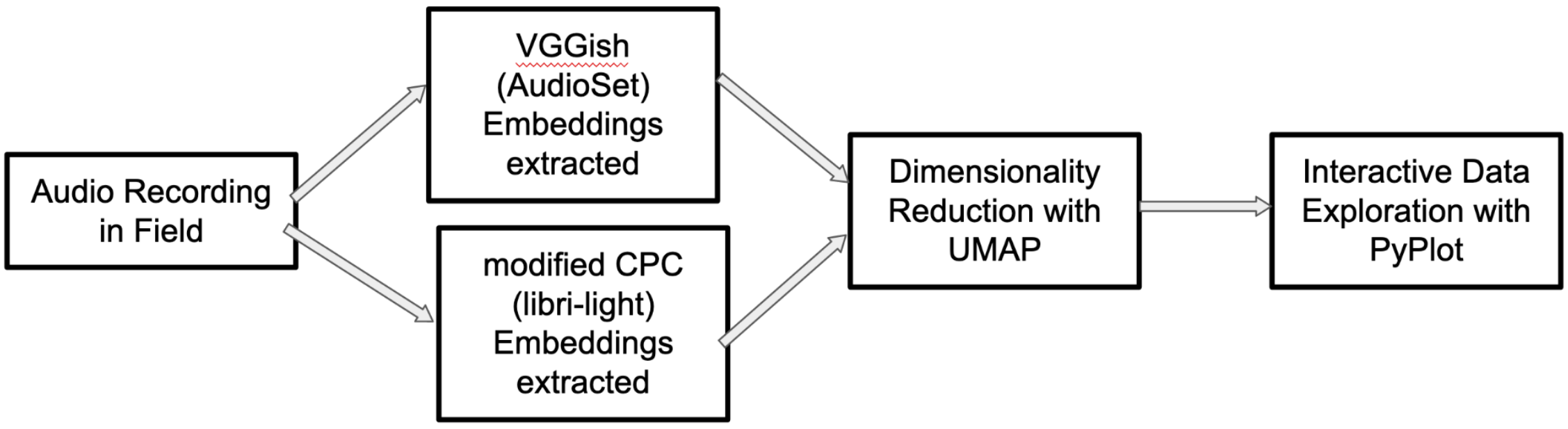

Fig. 2. Schematic of analysis workflow

**Frameworks**

1. VGGish using Audioset
2. Modified Contrastive Predictive Coding from S3PRL using Libri-light
3. PyPlot

**Steps**

1. 10 minute wav files resampled to match defaults in each model (16kHz). Default values were used for every other hyperparam in each framework.
2. Embeddings exported as pickle files, labelled with filename (date_time) (VGGish 0.96s segments, CPP 5s segments)
3. Dimensionality reduction using UMAP
4. Visualization using PyPlot & segment labels (date_Hour_Second) added as pop up

This creates an interactive data exploration pipeline. The UMAP can be explored and file name/ time cross-checked by auditioning manually.

## Results and Observations

Initial manual listening revealed relatively sparse events both above and below water.

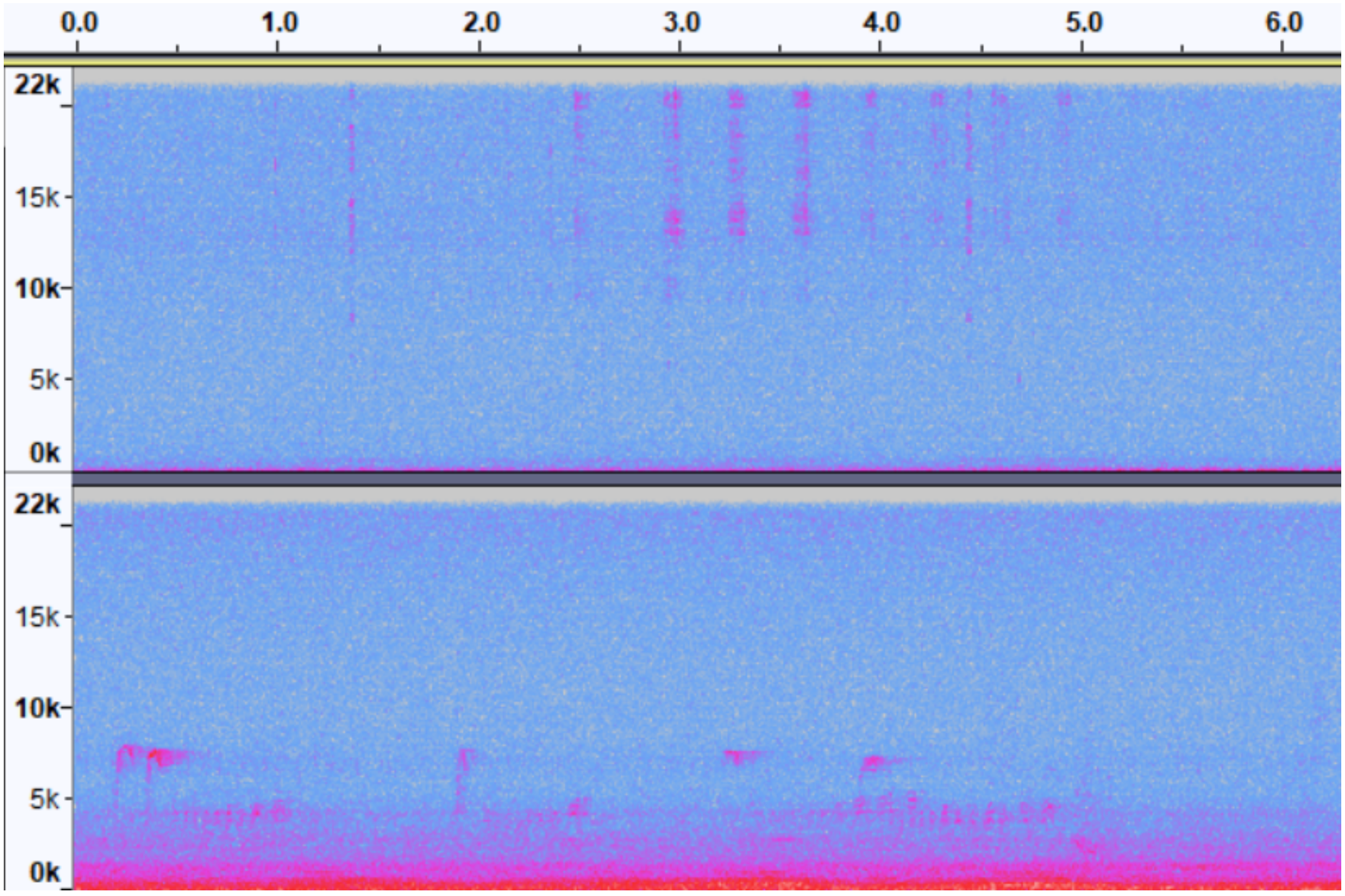

Fig 3. Example spectrogram of above (bottom) and below (top) soundscapes

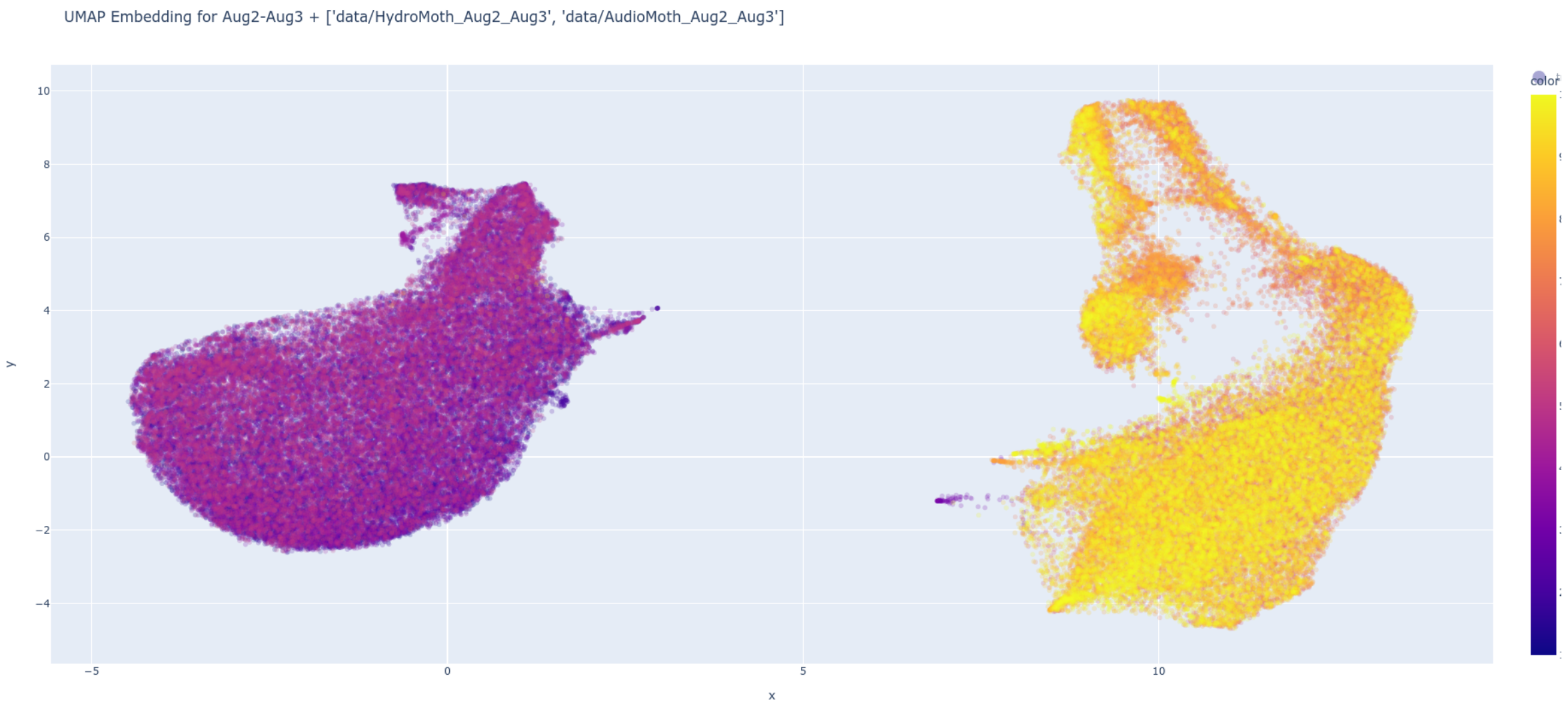

Fig 2. UMAP embeddings of Hydromoth (left) and Audiomoth (right)

## What we learned

### General

- Working with other curious people from different disciplines/ perspectives is really valuable
- Having serious time-constraints focuses the mind

### Hardware & set up

- Be careful when soldering sockets onto audiomoths
- Check deployment carefully: human error is often high - anticipating error and including redundancy in study design is important.
- How interesting the aquatic soundscape actually is! + the different way you have to think about exploring problems with aquatic sound
-

### Listening

- Apparent nocturnal patterns in vocalising water invertebrate (boatmen)

### Analysis

- Always use isolated development environment such as conda
- Specifics of different DL methods available especially the CPC (Liblight method)
- That it is possible to use embeddings without event detection
- Cool dynamic visualization is possible!
- The interactive plot speeds up the data exploration process
- How easy it is to switch models once you have a framework running
- To reign in software development ambitions when working to a deadline to ensure sufficient time to analyse and interpret outputs.

### Interpretation

- UMAP is an effective tool for data exploration - especially with pyplot
- An empty spectrogram (aka recording) with no events can also make an interesting data point (aka there is no such thing as noise)

**Link to Slides:** **https://docs.google.com/presentation/d/1rr-OJcB62uP0J3kgLM1itVutLiEhQYTuvZISupo2zww/edit?usp=sharing**

## Next Steps

**Ecological:**

- Do above water sound events penetrate aquatic soundscapes?
-

**Technical:**

- hardware: What is lost/ gained with high quality pre-amps underwater?

- Test efficacy of different embeddings generated from various state-of-the-art architectures that were pretrained Audioset vs Librispeech. The embeddings can be compared by retraining a classifier with the different embedding types. using a labelled dataset and classification task
- Transfer learning - retrain or fine-tune with ecological audio data
- Introduce audio-play back and/or spectrogram pop ups

## References

Riviere, M., Joulin, A., Mazaré, P.E. and Dupoux, E., 2020, May. Unsupervised pretraining transfers well across languages. In *ICASSP 2020-2020 IEEE International Conference on Acoustics, Speech and Signal Processing (ICASSP)* (pp. 7414-7418). IEEE.

Van der Lee, G.H., Desjonquères, C., Sueur, J., Kraak, M.H. and Verdonschot, P.F., 2020. Freshwater ecoacoustics: Listening to the ecological status of multi-stressed lowland waters. Ecological Indicators, 113, p.106252.

## Technical Resources

code created: https://github.com/BeckyHeath/AquaticEventDetection_FullStackBioacoustics
code used:
https://github.com/s3prl/s3prl/blob/master/s3prl/
https://github.com/ecila/models/tree/master/research/audioset/vggish

## SUPPLEMENTARY MATERIAL

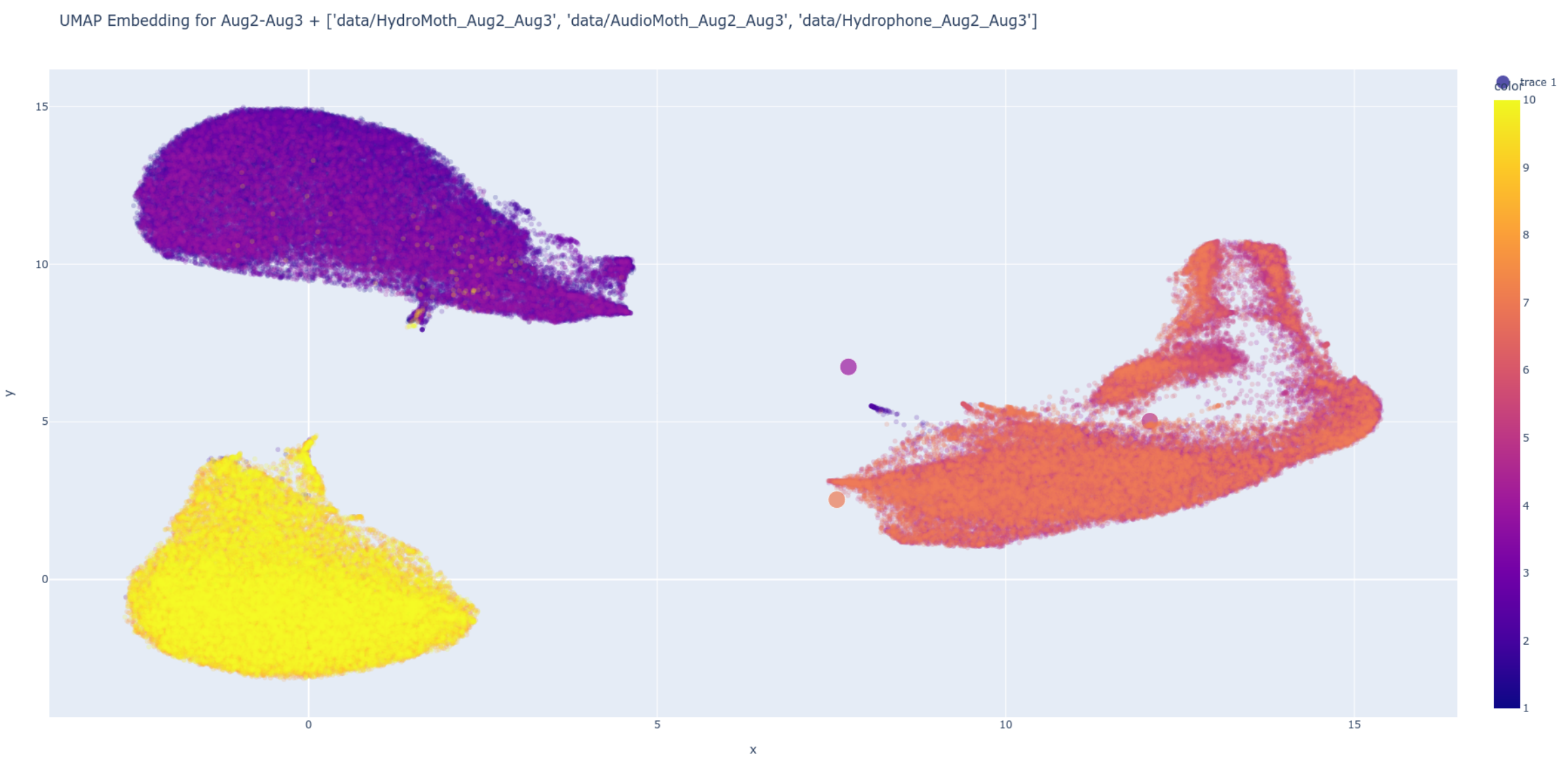

Fig SM1. Sanity Check. UMAP visualisation of HydroMoth (top left), Hydrophone (bottom left), and AudioMoth (right)

# Group-work overview: Group 6b

Group members: Avery Bick, Manjari Jain, Alison Johnston, Ricard Marxer, Julie Oswald, Carolyn Rosten

## Motivation/aims

Passive acoustic methods are widely used in monitoring programs for birds and other vocal taxa. While technology and pipelines to enable continuous recording are becoming available, sampling programs using some form of temporal subsampling (eg duty cycling of acoustic recordings) are still the norm. Design of temporal subsampling varies widely from study to study and may be based upon technical or practical constraints, data requirements, or be assigned randomly. Determining subsampling approaches based on the temporal and spatial acoustic behaviour of species of interest would maximise information gathered and enable better comparability between studies.

The aims of our work are to use passive acoustic data collected by the Sound of Norway project in 2022 to 1) design acoustic sampling regimes to determine bird biodiversity at multiple geographic locations in Norway, and 2) to use passive acoustic data to inform the timing of spot surveys conducted by ornithologists.

To accomplish these aims we will investigate the following questions:
<u>Aim 1 – Design acoustic sampling regimes</u>
Q1.1 How do the steepness and asymptote of species accumulation curves vary with different duty cycles?
Q1.2 How do species accumulation curves vary with location in Norway?
Q1.3 How do species accumulation curves vary on a larger geographic scale (Norway vs Ithaca, NY vs India)

<u>Aim 2: Informing the timing of spot surveys</u>
Q2.1 Is there a significant difference between trends in species richness determined by spot surveys vs passive acoustic data?
Q2.2 How sensitive is species detection to sampling week/day (ie how much are results affected by sampling at a time other than the peak in species richness)

## Hardware and deployment details

Over the last three years NINA (Norway Institute of Nature research) have been piloting the Sound of Norway project; a first of its kind nationwide real-time eco-acoustic monitoring network. The audio devices run continuously, transmitting acoustic data to NINA in near real-time over a mobile internet connection. Audio data is scanned for bird vocalisations using the BirdNet model and if present, identifies which species produced them. This allows monitoring of avian biodiversity on a scale and resolution that would not be possible using in-person point counts (the traditional method for collecting the same type of data). In 2021 NINA collected and analysed approximately 60,000 hours of data from over 40 sites across the network in the southern half of Norway. Data collection is under way on a similar scale this year covering the full latitudinal extent of Norway from Agder in the south to Finnmark in the north.

## Analysis methods used

For a preliminary analysis we used the 2022 Sound of Norway data, however, there were not sufficient data at enough sites to realise the aims of the project. We then used the 2021 Sound of Norway data for 21 sites, which have been analysed using BirdNet. The data were analysed in 3-second chunks of time, and we used any species identifications with at least 0.60 confidence from BirdNet.

The original recordings were continuous for 24 hours and we divided these up into different duty cycles and quantified the change in number of species recorded in two different ways:

a) The number of species accumulated over the whole year of recordings (Figure 1)
b) The species reported in a single week at a single site, as a proportion of all the species reported in that week at that site using 24 hour recordings.

We plan to produce a manuscript for submission to a peer reviewed journal once we have completed our data analyses. We have scheduled short Teams meetings every 2 weeks to help progress this project.

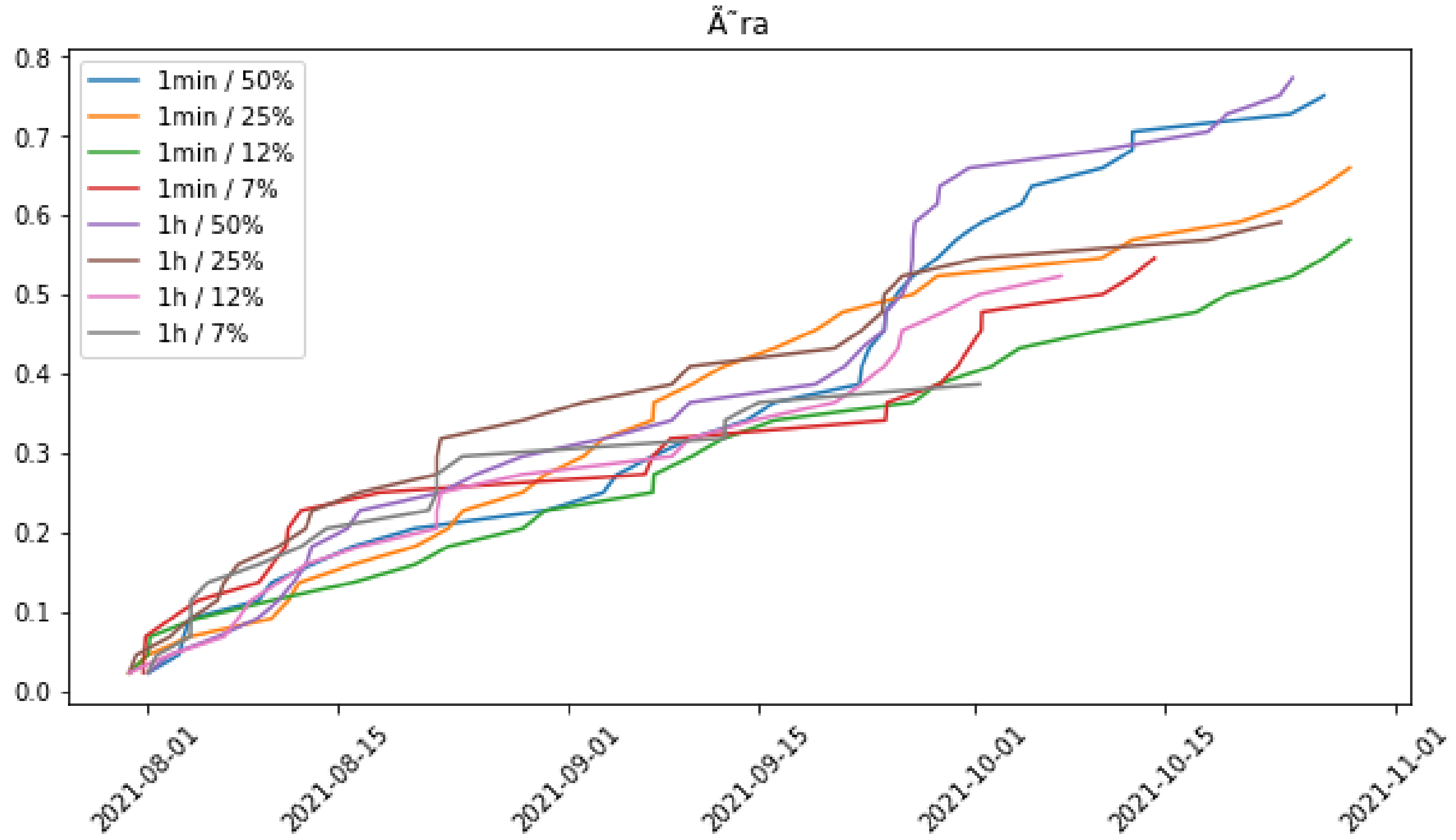

**Figure 1**. Accumulation of species across the whole year at a single site Åra. Each coloured line shows a different duty cycle. There is not a clear pattern of the 'best' duty cycle across all sites.

## What we learned

This was an incredibly useful week for all of us and we have taken a lot away from it. Some of the things we learned are as follows:

- How to put together the ‘full-stack’ in a bioacoustics project. ‘My eyes have been opened to the many different technology and analysis solutions available to solve our ecological questions’
- Who to contact with hardware, software, and analysis questions. The contacts made during this workshop have been amazing!
- How fun it is with people from different disciplines and share expertise (this is more a re-discovery)
- Another re-discovery – how important it is to step outside of our comfort zones and work with people from entirely different fields. The challenges we face are sometimes surprisingly similar
- How useful cross-disciplinary meet-ups are to advance the field in all of its spectrum
- How something as seemingly basic as deciding on a duty cycle could benefit from reflection and study and how analysis of new data enabledby recently developed devices and recording setups may shed some light on this question
- ‘More personally, due to my background in speech technology, I learned that perhaps some of the techniques and methodologies that we often recourse to may help several stages of the full-stack of bioacoustics. As examples we have the extensive use of microphone arrays, robust far-field audio processing techniques, and self-supervised learning methods.
- When you come up with a question that you think is niche, you will be surprised by how many other people are thinking about the same question.
- For machine learning, the dataset selections for training/test/validation are critical. The datasets are what create biases in the model. And the ‘right’ selection is not always obvious.
- The array of hardware available now. There is more than audiomoth!